\documentclass[twocolumn,floatfix,prb,aps,showpacs]{revtex4-2}
\usepackage{graphicx,amsmath,amssymb,color}
\usepackage{nicefrac}
\usepackage[titletoc,title]{appendix}
\usepackage[colorlinks,bookmarks=true,citecolor=blue,linkcolor=red,urlcolor=blue]{hyperref}

\begin{document}
	\title{Nature of the anomalous $4/13$ fractional quantum Hall effect in graphene}
	\author{Rakesh K. Dora and Ajit C. Balram}
	\affiliation{Institute of Mathematical Sciences, CIT Campus, Chennai, 600113, India}
	\affiliation{Homi Bhabha National Institute, Training School Complex, Anushaktinagar, Mumbai 400094, India}
	\date{\today}
	
	\begin{abstract}
	Extensive fractional quantum Hall effect (FQHE) has been observed in graphene-based materials. Some of the observed fractions are anomalous in that FQHE has not been established at these fractions in conventional GaAs systems. One such fraction is $4/13$, where incompressibility has recently been reported in graphene [Kumar \emph{et al.}, Nat. Commun. {\bf 9}, 2776 (2018)]. We propose a partonic wave function at $4/13$ and show it to be a viable candidate to describe the Coulomb ground state. Using the effective edge theory, we make predictions for experimentally measurable properties of the state.
	\end{abstract}	
	\maketitle
	
	The fractional quantum Hall effect (FQHE) appears when two-dimensional electrons are subjected to a strong perpendicular magnetic field and cooled to very low temperatures~\cite{Tsui82}. The magnetic field quenches the kinetic energy of the electrons leaving only the Coulomb interaction to determine the fate of the system. FQHE states possess topological order~\cite{Wen95} and thus FQHE systems provide a fertile platform to investigate strongly correlated topological phases of matter. Almost all of the experimentally observed FQHE states in the lowest Landau level (LLL) can be understood using Jain's composite fermion (CF) theory~\cite{Jain89}. The central postulate of the CF theory is that strongly interacting electrons turn into weakly interacting emergent particles called composite fermions (CFs), where a CF is a bound state of an electron and an even number, $2p$ ($p$ is a positive integer), of vortices. Due to the vortex binding, CFs feel a smaller effective magnetic field compared to the external magnetic field seen by the electrons. In the zeroth-order approximation, CFs are taken to be noninteracting and thus they form their own Landau-like levels called Lambda levels ($\Lambda$Ls). The electron filling factor $\nu$ is related to the CF filling factor $\nu^{*}$ as $\nu{=}\nu^{*}/(2p\nu ^{*}{\pm}1)$, where the ${+}$ (${-}$) sign denotes that the effective magnetic field sensed by the CFs is parallel (antiparallel) to the external magnetic field. Completely filling $n$ $\Lambda$Ls of CFs, i.e., $\nu^{*}{=}n$, can result in an incompressible state of electrons at $\nu{=}n/(2pn{\pm}1)$. The majority of FQHE states observed in the LLL lie in this $n/(2pn{\pm}1)$ Jain sequence. However, in the filling factor range $1/3{<}\nu{<}2/5$, there are fractions outside the Jain sequence such as $4/11$, $5/13$, $4/13$ where FQHE has been observed~\cite{Pan03, Pan15, Samkharadze15b, Kumar18a, Chung21}. In this Letter, we will consider FQHE at one of these fractions, namely $4/13$, a description of which lies beyond the ambit of the theory of free CFs.
	
	Pan~\emph{et al.}~\cite{Pan15} and Samkharadze~\emph{et al.}~\cite{Samkharadze15b} have established incompressibility at $4/11$ and $5/13$ in high-quality GaAs/AlGaAs semiconductor heterostructures. Signatures of FQHE at $\nu{=}4/13$ have been observed in some experiments~\cite{Pan03, Chung21} but incompressibility at this filling has not been established in GaAs-based materials. Extensive FQHE has also been observed in graphene-based materials~\cite{Neto09, Xu09, Bolotin09, Ghahari11, Dean11, Feldman12, Feldman13, Amet15, Kim19}. Suspended graphene samples have stronger electron-electron interaction than supported graphene samples and conventional semiconductor heterostructures and thus could potentially help realize delicate FQHE states. With this motivation, Kumar~\emph{et al.}~\cite{Kumar18a} designed a suspended graphene corbino disk sample and carried out conductance measurements on it. Along with states in the Jain sequence, they observed FQHE at $\nu{=}4/11$ and $4/13$ in the LLL. Furthermore, their transport measurements show activated behavior at $4/13$ thereby establishing incompressibility at this anomalous filling. The measured energy gap of the $4/13$ state is only $2{\%}$ of the $1/3$ gap indicating the fragile nature of the FQHE state at $4/13$. The spin polarization of the $4/13$ state has not been measured. We expect that the partially spin-polarized and spin-singlet $4/13$ states arise from the analogous CF states~\cite{Balram15, Balram16c}. Thus, we shall investigate only the fully polarized $4/13$ state here.
	
	Since the $4/13$ state is observed in the LLL, we expect the state to be well described by the CF theory in which CFs are interacting. The $\nu{=}4/13$ state of electrons is mapped onto the $\nu^{*}{=}4/3$ state of CFs carrying four vortices in a negative effective magnetic field. The $\nu^{*}{=}4/3$ state of CFs is obtained by filling the lowest $\Lambda$L (L$\Lambda$L) and forming a $1/3$ state of CFs in the second $\Lambda$L (S$\Lambda$L). An interesting question is what state do the CFs form at 1/3 filling in the S$\Lambda$L. Numerical calculations suggest that CFs do not form a 1/3 Laughlin state in the S$\Lambda$L~\cite{Mandal02, Mukherjee14}. This can be understood by noting that the interaction between CFs in their S$\Lambda$L in the presence of the filled L$\Lambda$L is dominated by repulsion in the relative angular momentum $m{=}3$ channel~\cite{Sitko96, Wojs00, Wojs04, Lee01, Lee02, Balram16c}, while the 1/3 Laughlin state is stabilized by interactions dominated by repulsion in the $m{=}1$ channel. Motivated by this observation, W\'ojs, Yi, and Quinn (WYQ) proposed that the interaction between CFs in their S$\Lambda$L can be simulated by the WYQ Hamiltonian $V_{m}{=}\delta_{m,3}$~\cite{Wojs00, Wojs04}, where $V_{m}$ is the Haldane pseudopotential~\cite{Haldane83} that denotes the interaction energy in the relative angular momentum $m$ channel. Exact diagonalization on the spherical geometry showed that the WYQ Hamiltonian has a uniform gapped ground state at $\nu{=}1/3$, referred to as the $1/3$ WYQ state that occurs at the Wen-Zee shift~\cite{Wen92} $\mathcal{S}{=}7$. This is to be contrasted with the 1/3 Laughlin state which occurs at $\mathcal{S}{=}3$. The shift characterizes the coupling of a quantum Hall state to background curvature~\cite{Wen92}. Two states that have different shifts carry different topological orders and thus the WYQ and Laughlin states are topologically distinct. Mukherjee and Mandal~\cite{Mukherjee15b} have shown that the exact LLL Coulomb ground state at $4/13$ is well represented by the $4/3$ state of CFs built from the 1/3 WYQ state in the S$\Lambda$L. An exact analytical closed-form expression of the 1/3 WYQ wave function is not known at present but progress has been made in constructing good candidate wave functions to describe the WYQ state~\cite{Das20}. 
	
	Numerical calculations in the past few years have called into question the proposition that the WYQ interaction leads to FQHE at 1/3.  Regnault~\emph{et al.}~\cite{Regnault16} studied the WYQ Hamiltonian at $\nu{=}1/3$ on the torus geometry and found that the ground state likely has spontaneously broken spatial symmetry, thereby suggesting that the WYQ interaction does not result in an FQHE ground state which is uniform. Furthermore, Misguich and Jolicoeur~\cite{Jolicoeur17a} observed that the charged and neutral excitation gaps of the $1/3$ WYQ state calculated in the spherical geometry extrapolate to a nonpositive value in the thermodynamic limit indicating that the $1/3$ WYQ state is gapless. Moreover, by studying the pair correlation function on the cylindrical geometry, they found that the 1/3 WYQ state forms a compressible bubble phase. These observations suggest that the WYQ Hamiltonian may not entirely capture the interaction between CFs in their S$\Lambda$L. To obtain an incompressible ground state at the WYQ shift one needs to supplement the WYQ Hamiltonian with other two-body pseudopotentials and/or include three or higher body interactions between the CFs~\cite{Balram17b}. 
	
	The above considerations show that a complete understanding of the mechanism of FQHE at $4/13$ in terms of CFs is lacking. In this work, without taking recourse to the CF theory, we directly propose a wave function for the $4/13$ state of electrons based on the parton theory~\cite{Jain89b}. Our ansatz gives a plausible description of the LLL Coulomb ground state at $4/13$ obtained in numerics. Based on the low-energy effective edge theory of the parton edge we make predictions that could be tested in experiments. We note that a parton wave function for the $4/11$ FQHE was recently constructed in Ref.~\cite{Balram21c}.

	\emph{Parton ansatz for $4/13$}. Jain proposed the parton theory~\cite{Jain89b} as a generalization of his CF theory to construct wave functions for incompressible state at all rational fillings. In the parton theory one divides an electron into $s$ different species, denoted by $\gamma{=}1,2,{\cdots},s$, of fictitious fermionic particles called partons. To obtain an incompressible state, each of the partons is placed in an integer quantum Hall effect (IQHE) state at filling factor $n_{\gamma}$ resulting in the state denoted as ``$n_{1},n_{2},n_{3}{\cdots}$'' and described by the wave function
	\begin{equation}
		\Psi_{\nu}^{n_{1},n_{2},n_{3}\cdots}= \mathcal{P}_{\rm LLL}\prod_{\gamma=1}^{s}\Phi_{n_{\gamma}}(\{z_{k}\}).
		\label{eq:general parton state}
	\end{equation}
	Here $\Phi_{n}$ is the Slater determinant wave function of the IQHE state with $n$-filled LLs of electrons, $z_{k}{=}x_{k}{-}iy_{k}$ is the two-dimensional coordinate of the $k^{\rm th}$ electron parametrized as a complex number, and $\mathcal{P}_{\rm LLL}$ projects the state into the LLL as is appropriate for the external magnetic field $B{\to}\infty$ limit. Antisymmetry of the electronic wave function in Eq.~\eqref{eq:general parton state} demands that $s$ is an odd integer. Partons can also ``feel" an effective magnetic field that is antiparallel to that of the electrons. We will denote such parton fillings by $\bar{n}{=}{-}n$ and the corresponding wave function as $\Phi_{\bar{n}}{=}[\Phi_{n}]^{*}$. The shift $\mathcal{S}$ of the state described by the wave function of Eq.~\eqref{eq:general parton state} is $\mathcal{S}=\sum_{\gamma{=}1}^{s}n_{\gamma}$.
	
	Identifying the coordinates of the different species of partons $z_{k}^{\gamma}$ with the electronic coordinate $z_{k}$, i.e., setting $z_{k}^{\gamma}{=}z_{k}~\forall\gamma$ in Eq.~\eqref{eq:general parton state}, glues the unphysical partons back into the electrons. The density of each parton species is the same as that of the parent electrons. Furthermore, each parton species is exposed to the same magnetic field as the underlying electrons. Thus the charge of the $\gamma$ parton $e_{\gamma}$ is related to the charge of the electron ${-}e$ as $e_{\gamma}{=}{-}e\nu/n_{\gamma}$. The constraint that the charges of the partons should add up to that of the electron relates the electronic and partonic fillings as $\nu^{-1}{=}\sum_{\gamma{=}1}^{s} n_{\gamma}^{{-}1}$. Parton states with a repeated factor of $n$ with $|n|{\geq}2$ host excitations that carry non-Abelian exchange statistics~\cite{Wen91}.
	
	Well-known FQHE states like Laughlin and Jain (CF) states can be reinterpreted as parton states. The Laughlin state~\cite{Laughlin83} at filling fraction $\nu{=}1/(2p{+}1)$, described by the wave function $\Psi^{\rm Laughlin}_{1/(2p+1)}=\Phi_{1}^{2p{+}1}$, is constructed using $(2p{+}1)$ species of partons and forming a $\nu{=}1$ IQH state for all of them. The Jain states at filling fraction $n/(2pn{\pm}1)$, described by the wave function $\Psi^{\rm Jain}_{n/(2pn{\pm}1)}=\mathcal{P}_{\rm LLL}\Phi_{\pm n}\Phi_{1}^{2p}$, are obtained from $(2p{+}1)$ partons, where $2p$ partons form $\nu{=}1$ IQH states and one parton forms a $\nu{=}{\pm}n$ IQH state. Using the parton theory we can construct FQHE states that lie beyond the purview of noninteracting CFs. Several such parton wave functions have recently been shown to be viable candidates to describe incompressible FQH states observed in a wide variety of settings such as in wide quantum wells, the second LL of GaAs, and in the LLs of graphene~\cite{Wu17, Balram18, Balram18a, Balram19, Bandyopadhyay18, Kim19, Faugno19, Balram20, Balram20a, Balram20b, Faugno20a, Faugno21, Balram21a, Balram21b, Balram21c}. 
	
	In this Letter, we propose the parton state denoted as ``$\bar{4}21^{3}$" and described by the wave function
	\begin{equation}
		\Psi^{\bar{4}21^{3}}_{4/13} = \mathcal{P}_{\rm LLL} [\Phi_{4}]^{*}\Phi_{2}\Phi^{3}_{1} \sim \frac{\Psi^{\rm Jain}_{4/7}\Psi^{\rm Jain}_{2/5}}{\Phi_{1}}
		\label{eq: parton_4_13_bar42111}
	\end{equation}
	to capture the $4/13$ ground state. The $\sim$ sign in the above equation indicates that the states on either side of the sign differ in how the LLL projection is implemented. We expect the topological phase that the wave function describes to be insensitive to such details~\cite{Balram15a, Balram16b}.  Only the wave function given on the rightmost side of Eq.~\eqref{eq: parton_4_13_bar42111} is readily amenable to an evaluation for very large system sizes since the constituent Jain states can be evaluated for hundreds of electrons using the Jain-Kamilla projection~\cite{Jain97, Davenport12, Balram15}. On the spherical geometry, this state occurs at the shift of $\mathcal{S}{=}1$. Interestingly, this value of the shift is identical to that of the 4/13 state built from the 1/3 WYQ state in the S$\Lambda$L~\cite{Mukherjee15b}. Moreover, $\mathcal{S}{=}1$ is the only known shift where the Coulomb ground state at $\nu{=}4/13$ in the LLL is consistently uniform for all systems for which numerical results are available~\cite{Mukherjee15b}.

	\emph{Numerical results}. All our numerical calculations are carried out on the Haldane sphere~\cite{Haldane83}. In this geometry, $N$ electrons move on the spherical surface in the presence of a radial magnetic flux of strength $2Qhc/e$ ($2Q$ is an integer) generated by a monopole placed at the center of the sphere.  The flux-particle relationship for a state on the sphere is written as $2Q{=}\nu^{{-}1}N{-}\mathcal{S}$, where the shift $\mathcal{S}$ is a quantum number characterizing the state~\cite{Wen92}. The IQH state of $n$ filled LLs occurs when $2Q{=}N/n{-}n$ and requires $N{\geq}n^2$ and for $N$ to be divisible by $n$.  Therefore, the $\bar{4}21^{3}$ state can be constructed on the sphere only when $N{=}16{+}4\alpha$, where $\alpha$ is a non-negative integer. Throughout this work, we will neglect the effects of LL mixing, finite width of the quantum well, and disorder. Under these assumptions, the FQHE physics in the LLL of graphene is identical to that in the LLL of GaAs~\cite{Balram15a}.

\begin{figure}[htpb]
		\begin{center}
			\includegraphics[width=0.47\textwidth,height=0.23\textwidth]{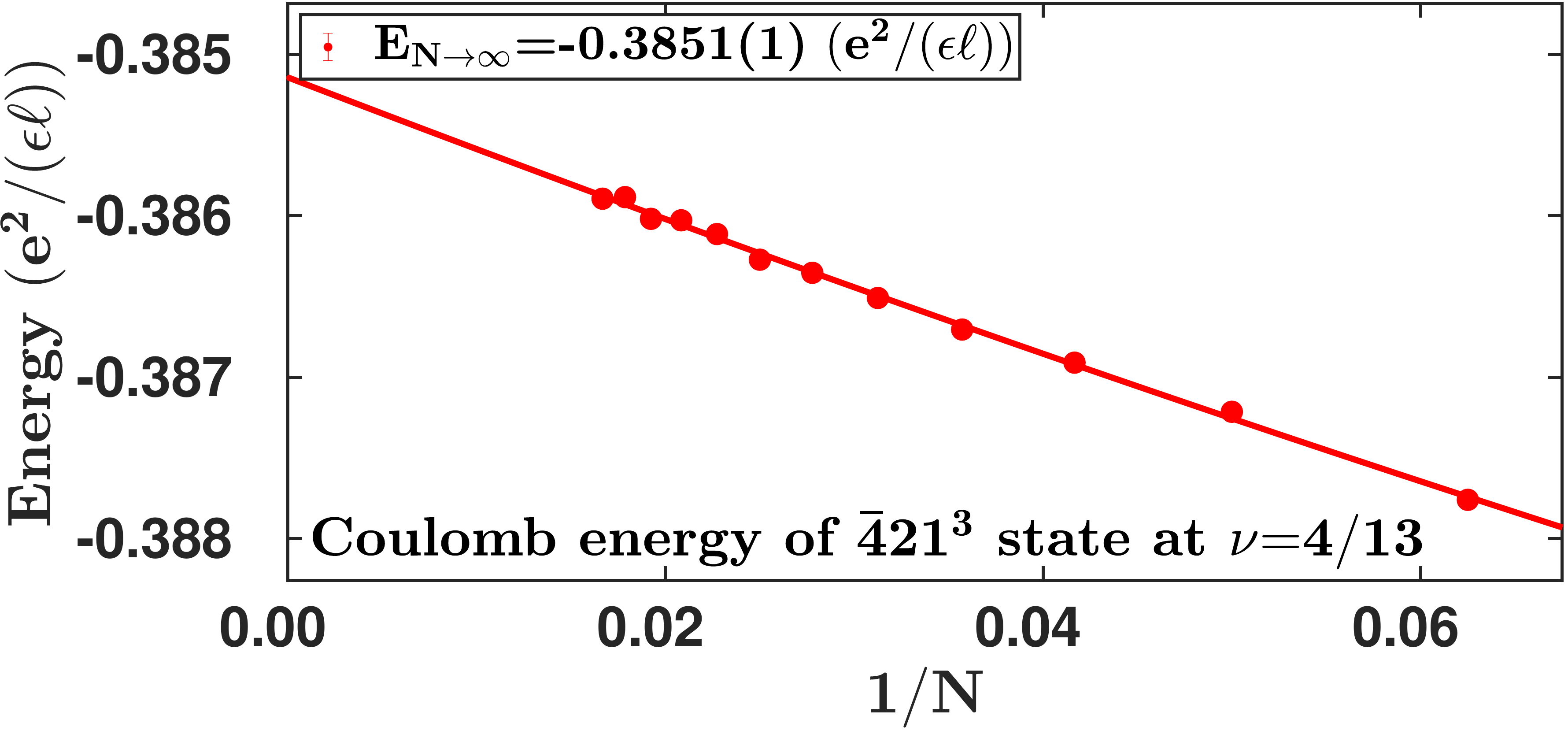}
			\caption{Thermodynamic extrapolation of the per-particle Coulomb energy of the $\bar{4}21^{3}$ state of Eq.~\eqref{eq: parton_4_13_bar42111} in the lowest Landau level as a linear function of $1/N$, where $N$ is the number of electrons, evaluated in the spherical geometry. Data is shown for $N{=}16$ to $60$ electrons.}
			\label{fig: energies_pp_bar42111_4_13}
		\end{center}
	\end{figure}

	The smallest system for which the $\bar{4}21^{3}$ state can be realized has $N{=}16$ electrons. This system has a dimension of over 81 billion that is beyond the reach of exact diagonalization. Thus, we cannot compare the $\bar{4}21^{3}$ ansatz against exact results. Nevertheless, in the LLL the method of CF diagonalization (CFD)~\cite{Mandal02} provides an almost-exact representation of the Coulomb ground states. Mukherjee and Mandal~\cite{Mukherjee15b} carried out extensive CFD calculations at $4/13$ and we shall compare the feasibility of the $\bar{4}21^{3}$ state against their CFD results. In the thermodynamic limit, including the contribution of the positively charged background, the density-corrected~\cite{Morf86} Coulomb energy of the CFD ground state at $\nu{=}4/13$ is ${-}0.3926(1)$ (the number in the parentheses denotes the error obtained from a linear extrapolation of the energy as a function of $1/N$), while that of the $\bar{4}21^{3}$ ansatz is ${-}0.3851(1)$ (see Fig.~\ref{fig: energies_pp_bar42111_4_13}). Although not as impressive as the agreement between the Laughlin/Jain wave functions and exact results, this agreement (within $2\%$) is on par with that between candidate states and exact results of other fragile states in the LLL~\cite{Balram21a, Balram21c}.

\begin{figure}[htpb]
		\begin{center}
			\includegraphics[width=0.47\textwidth,height=0.23\textwidth]{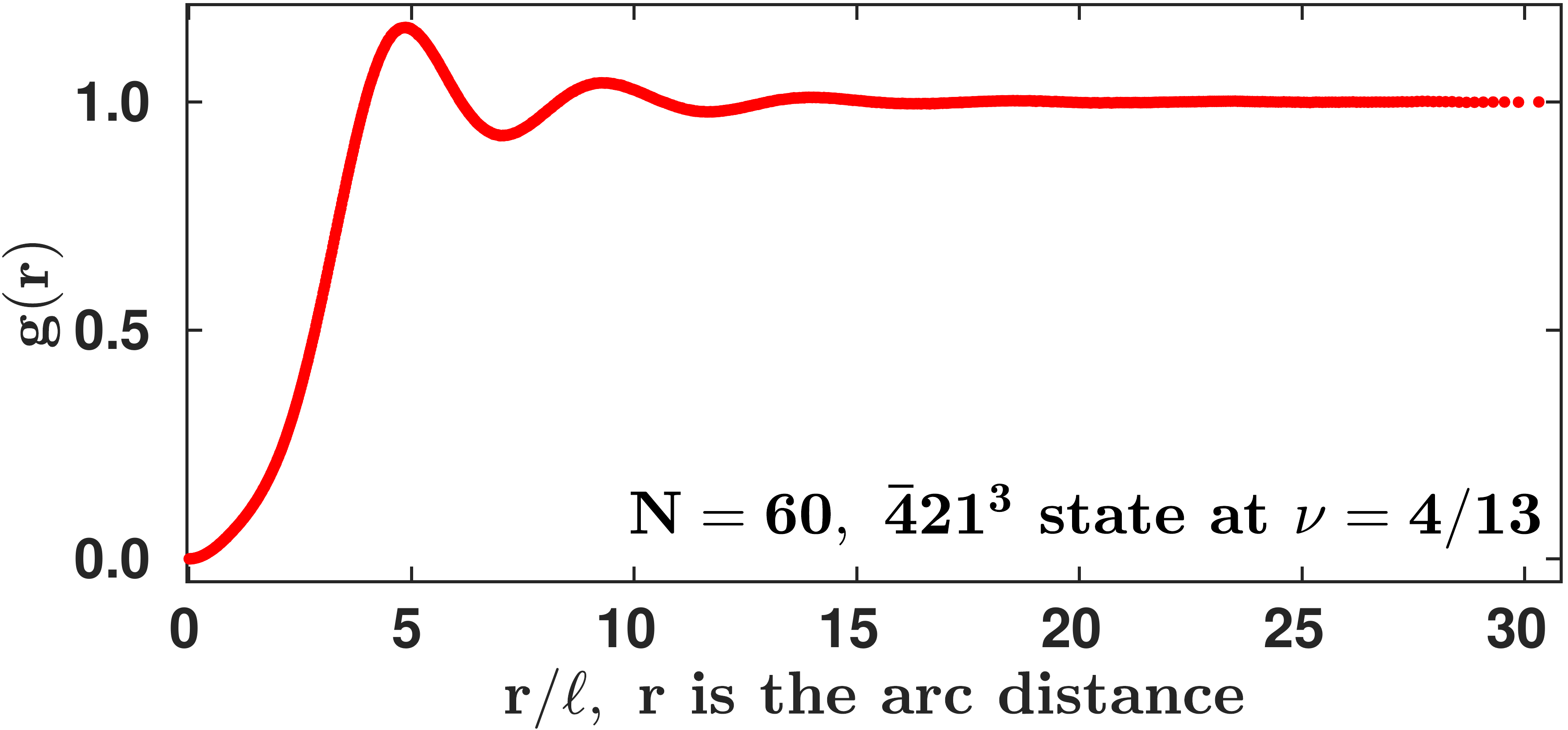}
			\caption{The pair-correlation function $g(r)$ of the $\bar{4}21^{3}$ state of Eq.~\eqref{eq: parton_4_13_bar42111} as a function of the arc distance $r$ on the sphere for $N{=}60$ electrons.}
			\label{fig: pair_correlations_bar42111_4_13}
		\end{center}
	\end{figure}

	As advertised above, the wave function given in Eq.~\eqref{eq: parton_4_13_bar42111} can be evaluated for large system sizes. As a proof of principle, in Fig.~\ref{fig: pair_correlations_bar42111_4_13} we show the pair-correlation function $g(\vec{r}){=}[1/(\rho N)]\left\langle\sum_{i\neq j}\delta^{(2)}\left(\vec{r}{-}\left[\vec{r}_{i}{-}\vec{r}_{j}\right]\right)\right\rangle$, where $\rho$ is the density, of the $\bar{4}21^{3}$ state for $N{=}60$ electrons [due to spherical symmetry $g(\vec{r}){\equiv}g(r)$]. The pair-correlation function is similar to that of other incompressible FQH fluids~\cite{Balram15b, Balram17} in that it has oscillations at short-to-intermediate distances and goes to unity at large distances, i.e., $g(r){\to}1$ as $r{\to}\infty$. The absence of a ``shoulder"-like feature at intermediate distances, which typically exists in non-Abelian states~\cite{Read99, Hutasoit16, Balram20a}, is in accordance with the $\bar{4}21^{3}$ state being Abelian.

\begin{figure}[htpb]
		\begin{center}
			\includegraphics[width=0.47\textwidth,height=0.23\textwidth]{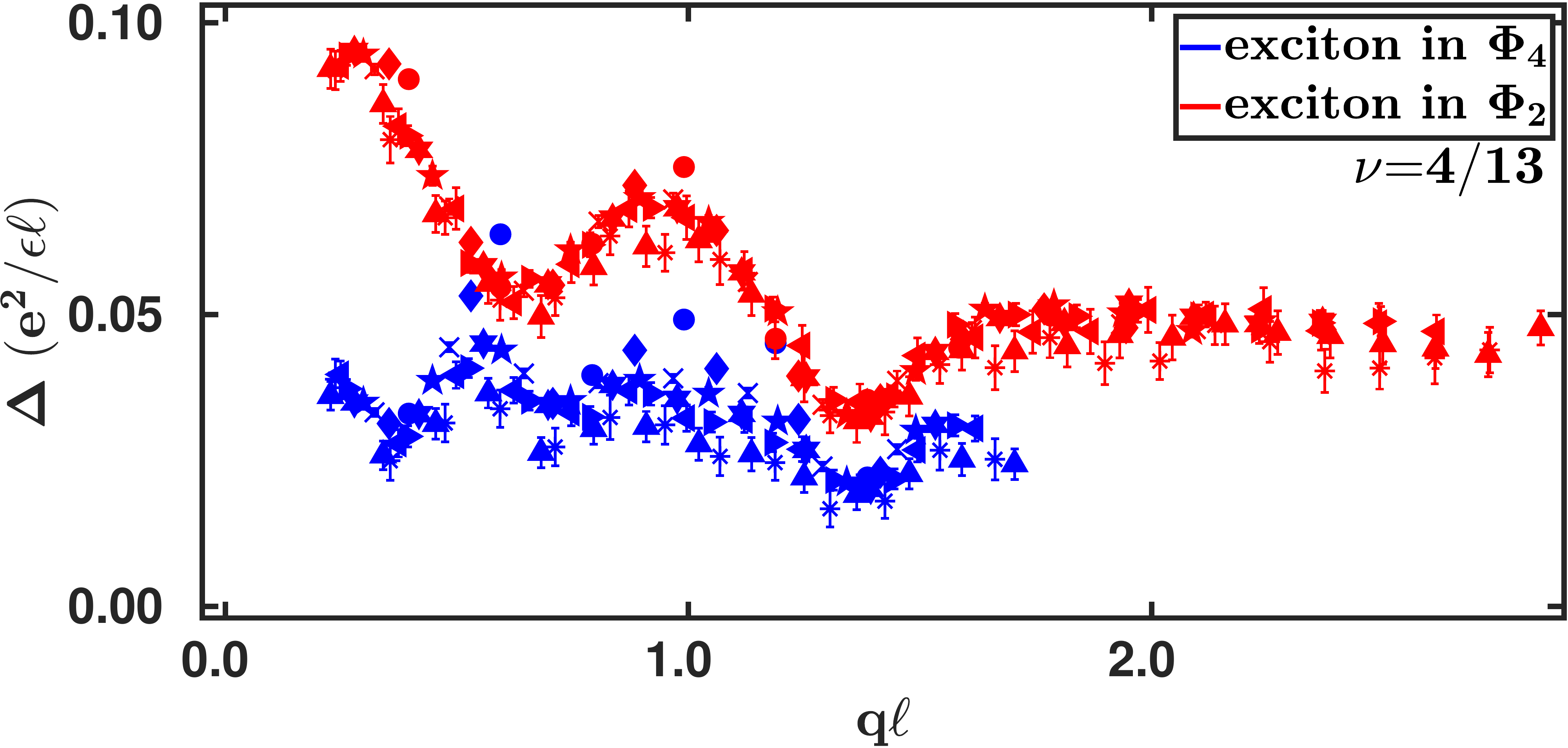}
			\caption{Lowest Landau level Coulomb gaps $\Delta$ of the low-(blue) and high-energy (red) parton-based collective modes at $\nu{=}4/13$ (see text). Data points for different system sizes from $N{=}16$ to $48$ electrons are plotted with different symbols.}
			\label{fig: excitons_bar42111_4_13}
		\end{center}
	\end{figure}

	Next, we look at the gaps of the $4/13$ state. The lowest-energy neutral excitation is obtained by creating a particle-hole pair in the $\Phi_{4}$ factor since the corresponding parton carries the smallest charge of magnitude $e/13$. This results in the wave function $\Psi^{{\rm mode-a},L}_{4/13} {\sim}  \Psi^{{\rm CFE},L}_{4/7} \Psi^{\rm Jain}_{2/5}/\Phi_{1}$, where $\Psi^{{\rm CFE},L}_{n/(2n\pm 1)}$ is the CF-exciton (CFE) state at $\nu{=}n/(2n{\pm}1)$~\cite{Kamilla96b} with orbital angular momentum $L=2,3,{\cdots},(N/n{-}n){+}2n{-}1$~\cite{Balram16d}.  The dispersion of this mode as a function of the linear momentum $q\ell{=}L/\sqrt{Q}$ is shown in Fig.~\ref{fig: excitons_bar42111_4_13}. The mode has an anomalous minimum in energy at wave number $q\ell{\approx}0.25$ consistent with the results of Mukherjee and Mandal~\cite{Mukherjee15} who came to the same conclusion using the Girvin-MacDonald-Platzman (GMP) density-mode ansatz~\cite{Girvin85, Girvin86}. Very recently, it has been noticed that the parton theory can also describe certain very-high-energy excitations of FQH fluids~\cite{Balram21d}. The parton ansatz of Eq.~\eqref{eq: parton_4_13_bar42111} naturally suggests the wave function $\Psi^{{\rm mode-b},L}_{4/13} {\sim}\Psi^{\rm Jain}_{4/7} \Psi^{{\rm CFE},L}_{2/5}/\Phi_{1}$, which describes a high-energy neutral collective mode at $4/13$ that also starts from $L{=}2$. For completeness, we also show its dispersion in Fig.~\ref{fig: excitons_bar42111_4_13}. The two modes $\Psi^{{\rm mode-a},L}_{4/13}$ and $\Psi^{{\rm mode-b},L}_{4/13}$ carry opposite chiralities~\cite{Liou19} since the partons hosting the excitation see effective magnetic fields in the opposite directions~\cite{Balram21d}. In the large wave number limit, i.e., $q\ell{\to}\infty$, the dispersion of both the modes flattens out since the constituent CF particle and CF hole of the CFE are far from each other and thus the interaction between them is negligible.  We have not been able to get a reliable estimate of the thermodynamic limit of the transport gap at $4/13$, the $q\ell{\to}\infty$ gap of $\Psi^{{\rm mode-a}, L}_{4/13}$, as its dispersion has not flattened out for the systems accessible to us.

	We mention here that recently the collective modes of the $\nu^{*}{=}1{+}1/3$ Laughlin-based $4/13$ state~\cite{Lopez04} (that is unlikely to be stabilized by Coulomb interaction in the LLL) have been studied by Wang and Yang~\cite{Wang22} using the dynamical structure factor. They found two peaks in the dynamical structure factor indicating the presence of two density modes. Here we provide parton wave functions to describe these modes. The ground state wave function for this conventional $4/13$ state~\cite{Balram15} can be factorized as $\Psi_{4/13} {\sim} \Psi_{4/5}{\times} \Psi^{\rm Laughlin}_{1/2}$ ($\Psi_{4/5}{=}\mathcal{P}_{\rm LLL}[\Psi_{4/3}]^{*}\Phi^{2}_{1}$ is topologically equivalent to the hole conjugate of the $1/5$ Laughlin state~\cite{Balram16c}). The low- and high-energy modes can respectively be described by the wave functions $\Psi^{{\rm mode-1},L}_{4/13} {\sim} \Psi^{{\rm CFE},L}_{4/5}{\times} \Psi^{\rm Laughlin}_{1/2}$ and $\Psi^{{\rm mode-2},L}_{4/13} {\sim} \Psi_{4/5}{\times} \Psi^{{\rm CFE},L}_{1/2}$ (alternatively, one could use the GMP ansatz instead of the CFE).  As in the $\bar{4}21^{3}$ state, this construction suggests that the two modes of the conventional $4/13$ state carry opposite chiralities. The clustering/vanishing properties of the two modes can be determined by these wave functions. When two electrons, with relative separation $r$, are brought close together, $|\Psi^{{\rm mode-1},L}_{4/13}|{\sim}r^{3}$ while $|\Psi^{{\rm mode-2},L}_{4/13}|{\sim}r$. These observations are consistent with the results of Ref.~\cite{Wang22}. 
	
	\emph{Effective theory of the $\bar{4}21^{3}$ edge}. We now turn to the universal properties of the $\bar{4}21^{3}$ state that can be obtained from the low-energy effective theory of its edge. The effective edge theory of the $\bar{4}21^{3}$ state can be derived following standard techniques which have been discussed in detail previously~\cite{Wen91b, Wen92b, Wen95, Moore98, Balram20a}, so here we will just present the main results. The topological properties of the $\bar{4}21^{3}$ state are encoded in the $K$ matrix, charge vector $t$, and the spin vector $\mathfrak{s}$, which are given by
	\begin{equation}
		K{=}
		\begin{pmatrix}
			{-}2 & {-}1 & {-}1 & 0 & 1 \\
			{-}1 &  {-}2 & {-}1 & 0 & 1 \\
			{-}1 & {-}1 & {-}2 & 0 & 1 \\
			0 &  0 & 0 &2 & {-}1 \\
			1 & 1 &  1 & {-}1 & 3\\
		\end{pmatrix},~
		t{=}\begin{pmatrix}
			0\\
			0\\
			0\\
			0\\
			1\\
		\end{pmatrix},~{\rm and}~
		\mathfrak{s}{=}\begin{pmatrix}
			3\\
			2\\
			1\\
			{-}1\\
			{-}1/2\\
		\end{pmatrix}.
	\end{equation}
	This results in a filling fraction of $\nu {=}  t^{\rm T}{\cdot} K^{{-}1} {\cdot} t{=} 4/13$, and shift $\mathcal{S}{=}(2/\nu) \left( t^{\rm T}{\cdot} K^{{-}1} {\cdot} \mathfrak{s} \right){=}1$, consistent with the values inferred from the microscopic wave function given in Eq.~\eqref{eq: parton_4_13_bar42111}. The degeneracy of the ground state on a torus $|{\rm det}(K)| {=}26$. An interesting aspect to note is that the $\bar{4}21^{3}$ state is a single-component Abelian state at $\nu{=}a/b$ (with $a,b$ coprime positive integers), which has a ground state degeneracy on the torus that is greater than the denominator $b$ (see Refs.~\cite{Balram20, Balram20b} for other examples of such states). This $K$ matrix has three negative and two positive eigenvalues, which indicates that the $\bar{4}21^{3}$ state has two downstream and three upstream edge modes resulting in a chiral central charge of $c_{-}{=}{-}1$.  An intuitive way to see the presence of five edge modes is as follows: at the mean-field level the parton theory results in a total of nine-edge states four from the factor $\Phi_{\bar{4}}$, two from $\Phi_{2}$, and one each from each factor of $\Phi_{1}$.  However, these edge states are not all independent since the density variations of all the five partons must be identified, which results in precisely four constraints and thereby leads to exactly five independent edge states.
	
	\emph{Discussion}. We now discuss experimentally testable properties of the $\bar{4}21^{3}$ state that can reveal its underlying topological order. The fundamental or smallest charged quasiparticle is obtained by creating a particle in the $\Phi_{\bar{4}}$ factor and carries a charge of ${-}e/13$. Quasiparticles with larger magnitude charges can be produced by creating a particle in the $\Phi_{2}$ or $\Phi_{1}$ factors. All these excitations of the $\bar{4}21^{3}$ state possess Abelian braid statistics.
	
Heat transport measurements at $4/13$ could help identify the underlying partonic order. In the past few years, thermal Hall measurements have been carried out in both GaAs~\cite{Banerjee17, Banerjee17b} and in graphene~\cite{Srivastav19, Srivastav21}.  \emph{Assuming} a full equilibration of the edge channels, the thermal Hall conductance $\kappa_{xy}$ of the $\bar{4}21^{3}$ state, at temperatures much lower than the gap, is $\kappa_{xy}{=}c_{-}[\pi^2 k^{2}_{B}/(3h)T]{=}{-}1[\pi^2 k^{2}_{B}/(3h)T]$. At the lowest accessible temperatures, where thermal equilibration is poor, the thermal Hall conductance of the proposed state is therefore expected to be larger than $[\pi^2 k^{2}_{B}/(3h)T]$~\cite{Banerjee17b, Ravi} (current thermal Hall conductance measurements are not sensitive to the sign of $\kappa_{xy}$). Hence experimental observation of the thermal Hall conductance decreasing to its lowest value of $[\pi^2 k^{2}_{B}/(3h)T]$ with increasing temperature (with the temperature still much lower than the gap) would validate the proposed edge structure of the $4/13$ state. The Hall viscosity $\eta_{H}$ of the $\bar{4}21^{3}$ state is also anticipated to be quantized~\cite{Read09} as $\eta_{H}{=}(4/13)\hbar \mathcal{S}/(8\pi\ell^{2})$, where $\mathcal{S}{=}1$ for the $\bar{4}21^{3}$ ansatz.

Another experimentally measurable quantity is the tunneling exponent of quasiparticles across the FQH edges.  The tunneling exponent corresponding to an FQH state having all the edge modes of the same chirality is universal, i.e., for fully chiral states the exponent does not depend on the specific details of the interaction between the edge modes. In such a scenario the tunneling exponent is related to the bulk topological order, in particular to the quasiparticle statistics~\cite{Wen95}.  Since the $\bar{4}21^{3}$ harbors both upstream- and downstream-propagating modes its tunneling exponents will not be quantized to a universal value.  For this reason, we have not calculated the tunneling exponents of the $\bar{4}21^{3}$ state.
	
	An interpretation of our parton state as the $\nu^{*}{=}1{+}1/3$ state of CFs carrying four vortices in a negative effective magnetic field could then, by particle-hole conjugation of CFs in their S$\Lambda$L~\cite{Balram17b}, also explain the $5/17$ FQHE which corresponds to $\nu^{*}{=}1{+}2/3$ of CFs carrying four vortices in a negative effective magnetic field. We have not been able to connect our $4/13$ parton ansatz to a $1/3$ state of interacting CFs in their S$\Lambda$L. 
	
	The $\bar{4}21^{3}$ naturally suggests two families of parton states, namely $\bar{4}n1^{3}$ and $\bar{n}21^{3}$, whose properties we briefly mention next. The $\bar{4}n1^{3}$ sequence produces states at $\nu{=}4n/(11n{+}4)$ with shift $\mathcal{S}{=}n{-}1$. On the other hand, the $\bar{n}21^{3}$ sequence produces states at $\nu{=}2n/(7n{-}2)$ with shift $\mathcal{S}{=}5{-}n$. Signatures of FQHE have been observed at some members of the above sequences (aside from some fractions which also lie in the Jain sequence, for which the Jain states would be more suitable candidates in the LLL), such as at $\nu{=}3/10$~\cite{Pan03} which could be described by the $\bar{6}21^{3}$. However, definitive evidence of incompressibility at these fractions is lacking.

	\textit{Acknowledgements} - We acknowledge useful discussions with J. K. Jain. Computational portions of this research work were conducted using the Nandadevi supercomputer, which is maintained and supported by the Institute of Mathematical Science's High-Performance Computing Center. We thank the Science and Engineering Research Board (SERB) of the Department of Science and Technology (DST) for funding support via Start-up Grant No. SRG/2020/000154.

	\bibliography{biblio_fqhe}

\begin{thebibliography}{78}
\expandafter\ifx\csname natexlab\endcsname\relax\def\natexlab#1{#1}\fi
\expandafter\ifx\csname bibnamefont\endcsname\relax
  \def\bibnamefont#1{#1}\fi
\expandafter\ifx\csname bibfnamefont\endcsname\relax
  \def\bibfnamefont#1{#1}\fi
\expandafter\ifx\csname citenamefont\endcsname\relax
  \def\citenamefont#1{#1}\fi
\expandafter\ifx\csname url\endcsname\relax
  \def\url#1{\texttt{#1}}\fi
\expandafter\ifx\csname urlprefix\endcsname\relax\def\urlprefix{URL }\fi
\providecommand{\bibinfo}[2]{#2}
\providecommand{\eprint}[2][]{\url{#2}}

\bibitem[{\citenamefont{Tsui et~al.}(1982)\citenamefont{Tsui, Stormer, and
  Gossard}}]{Tsui82}
\bibinfo{author}{\bibfnamefont{D.~C.} \bibnamefont{Tsui}},
  \bibinfo{author}{\bibfnamefont{H.~L.} \bibnamefont{Stormer}},
  \bibnamefont{and} \bibinfo{author}{\bibfnamefont{A.~C.}
  \bibnamefont{Gossard}}, \bibinfo{journal}{Phys. Rev. Lett.}
  \textbf{\bibinfo{volume}{48}}, \bibinfo{pages}{1559} (\bibinfo{year}{1982}),
  \urlprefix\url{http://link.aps.org/doi/10.1103/PhysRevLett.48.1559}.

\bibitem[{\citenamefont{Wen}(1995)}]{Wen95}
\bibinfo{author}{\bibfnamefont{X.-G.} \bibnamefont{Wen}},
  \bibinfo{journal}{Advances in Physics} \textbf{\bibinfo{volume}{44}},
  \bibinfo{pages}{405} (\bibinfo{year}{1995}),
  \eprint{http://www.tandfonline.com/doi/pdf/10.1080/00018739500101566},
  \urlprefix\url{http://www.tandfonline.com/doi/abs/10.1080/00018739500101566}.

\bibitem[{\citenamefont{Jain}(1989{\natexlab{a}})}]{Jain89}
\bibinfo{author}{\bibfnamefont{J.~K.} \bibnamefont{Jain}},
  \bibinfo{journal}{Phys. Rev. Lett.} \textbf{\bibinfo{volume}{63}},
  \bibinfo{pages}{199} (\bibinfo{year}{1989}{\natexlab{a}}),
  \urlprefix\url{http://link.aps.org/doi/10.1103/PhysRevLett.63.199}.

\bibitem[{\citenamefont{Pan et~al.}(2003)\citenamefont{Pan, Stormer, Tsui,
  Pfeiffer, Baldwin, and West}}]{Pan03}
\bibinfo{author}{\bibfnamefont{W.}~\bibnamefont{Pan}},
  \bibinfo{author}{\bibfnamefont{H.~L.} \bibnamefont{Stormer}},
  \bibinfo{author}{\bibfnamefont{D.~C.} \bibnamefont{Tsui}},
  \bibinfo{author}{\bibfnamefont{L.~N.} \bibnamefont{Pfeiffer}},
  \bibinfo{author}{\bibfnamefont{K.~W.} \bibnamefont{Baldwin}},
  \bibnamefont{and} \bibinfo{author}{\bibfnamefont{K.~W.} \bibnamefont{West}},
  \bibinfo{journal}{Phys. Rev. Lett.} \textbf{\bibinfo{volume}{90}},
  \bibinfo{pages}{016801} (\bibinfo{year}{2003}),
  \urlprefix\url{http://link.aps.org/doi/10.1103/PhysRevLett.90.016801}.

\bibitem[{\citenamefont{Pan et~al.}(2015)\citenamefont{Pan, Baldwin, West,
  Pfeiffer, and Tsui}}]{Pan15}
\bibinfo{author}{\bibfnamefont{W.}~\bibnamefont{Pan}},
  \bibinfo{author}{\bibfnamefont{K.~W.} \bibnamefont{Baldwin}},
  \bibinfo{author}{\bibfnamefont{K.~W.} \bibnamefont{West}},
  \bibinfo{author}{\bibfnamefont{L.~N.} \bibnamefont{Pfeiffer}},
  \bibnamefont{and} \bibinfo{author}{\bibfnamefont{D.~C.} \bibnamefont{Tsui}},
  \bibinfo{journal}{Phys. Rev. B} \textbf{\bibinfo{volume}{91}},
  \bibinfo{pages}{041301} (\bibinfo{year}{2015}),
  \urlprefix\url{http://link.aps.org/doi/10.1103/PhysRevB.91.041301}.

\bibitem[{\citenamefont{Samkharadze et~al.}(2015)\citenamefont{Samkharadze,
  Arnold, Pfeiffer, West, and Cs\'athy}}]{Samkharadze15b}
\bibinfo{author}{\bibfnamefont{N.}~\bibnamefont{Samkharadze}},
  \bibinfo{author}{\bibfnamefont{I.}~\bibnamefont{Arnold}},
  \bibinfo{author}{\bibfnamefont{L.~N.} \bibnamefont{Pfeiffer}},
  \bibinfo{author}{\bibfnamefont{K.~W.} \bibnamefont{West}}, \bibnamefont{and}
  \bibinfo{author}{\bibfnamefont{G.~A.} \bibnamefont{Cs\'athy}},
  \bibinfo{journal}{Phys. Rev. B} \textbf{\bibinfo{volume}{91}},
  \bibinfo{pages}{081109} (\bibinfo{year}{2015}),
  \urlprefix\url{http://link.aps.org/doi/10.1103/PhysRevB.91.081109}.

\bibitem[{\citenamefont{Kumar et~al.}(2018)\citenamefont{Kumar, Laitinen, and
  Hakonen}}]{Kumar18a}
\bibinfo{author}{\bibfnamefont{M.}~\bibnamefont{Kumar}},
  \bibinfo{author}{\bibfnamefont{A.}~\bibnamefont{Laitinen}}, \bibnamefont{and}
  \bibinfo{author}{\bibfnamefont{P.}~\bibnamefont{Hakonen}},
  \bibinfo{journal}{Nature Communications} \textbf{\bibinfo{volume}{9}},
  \bibinfo{pages}{2776} (\bibinfo{year}{2018}), ISSN \bibinfo{issn}{2041-1723},
  \urlprefix\url{https://doi.org/10.1038/s41467-018-05094-8}.

\bibitem[{\citenamefont{Chung et~al.}(2021)\citenamefont{Chung,
  Villegas~Rosales, Baldwin, Madathil, West, Shayegan, and Pfeiffer}}]{Chung21}
\bibinfo{author}{\bibfnamefont{Y.~J.} \bibnamefont{Chung}},
  \bibinfo{author}{\bibfnamefont{K.~A.} \bibnamefont{Villegas~Rosales}},
  \bibinfo{author}{\bibfnamefont{K.~W.} \bibnamefont{Baldwin}},
  \bibinfo{author}{\bibfnamefont{P.~T.} \bibnamefont{Madathil}},
  \bibinfo{author}{\bibfnamefont{K.~W.} \bibnamefont{West}},
  \bibinfo{author}{\bibfnamefont{M.}~\bibnamefont{Shayegan}}, \bibnamefont{and}
  \bibinfo{author}{\bibfnamefont{L.~N.} \bibnamefont{Pfeiffer}},
  \bibinfo{journal}{Nature Materials}  (\bibinfo{year}{2021}), ISSN
  \bibinfo{issn}{1476-4660},
  \urlprefix\url{https://doi.org/10.1038/s41563-021-00942-3}.

\bibitem[{\citenamefont{Castro~Neto et~al.}(2009)\citenamefont{Castro~Neto,
  Guinea, Peres, Novoselov, and Geim}}]{Neto09}
\bibinfo{author}{\bibfnamefont{A.~H.} \bibnamefont{Castro~Neto}},
  \bibinfo{author}{\bibfnamefont{F.}~\bibnamefont{Guinea}},
  \bibinfo{author}{\bibfnamefont{N.~M.~R.} \bibnamefont{Peres}},
  \bibinfo{author}{\bibfnamefont{K.~S.} \bibnamefont{Novoselov}},
  \bibnamefont{and} \bibinfo{author}{\bibfnamefont{A.~K.} \bibnamefont{Geim}},
  \bibinfo{journal}{Rev. Mod. Phys.} \textbf{\bibinfo{volume}{81}},
  \bibinfo{pages}{109} (\bibinfo{year}{2009}),
  \urlprefix\url{http://link.aps.org/doi/10.1103/RevModPhys.81.109}.

\bibitem[{\citenamefont{Du et~al.}(2009)\citenamefont{Du, Skachko, Duerr,
  Luican, and Andrei}}]{Xu09}
\bibinfo{author}{\bibfnamefont{X.}~\bibnamefont{Du}},
  \bibinfo{author}{\bibfnamefont{I.}~\bibnamefont{Skachko}},
  \bibinfo{author}{\bibfnamefont{F.}~\bibnamefont{Duerr}},
  \bibinfo{author}{\bibfnamefont{A.}~\bibnamefont{Luican}}, \bibnamefont{and}
  \bibinfo{author}{\bibfnamefont{E.~Y.} \bibnamefont{Andrei}},
  \bibinfo{journal}{Nature} \textbf{\bibinfo{volume}{462}},
  \bibinfo{pages}{192} (\bibinfo{year}{2009}).

\bibitem[{\citenamefont{Bolotin et~al.}(2009)\citenamefont{Bolotin, Ghahari,
  Shulman, Stormer, and Kim}}]{Bolotin09}
\bibinfo{author}{\bibfnamefont{K.}~\bibnamefont{Bolotin}},
  \bibinfo{author}{\bibfnamefont{F.}~\bibnamefont{Ghahari}},
  \bibinfo{author}{\bibfnamefont{M.~D.} \bibnamefont{Shulman}},
  \bibinfo{author}{\bibfnamefont{H.}~\bibnamefont{Stormer}}, \bibnamefont{and}
  \bibinfo{author}{\bibfnamefont{P.}~\bibnamefont{Kim}},
  \bibinfo{journal}{Nature} \textbf{\bibinfo{volume}{462}},
  \bibinfo{pages}{196} (\bibinfo{year}{2009}).

\bibitem[{\citenamefont{Ghahari et~al.}(2011)\citenamefont{Ghahari, Zhao,
  Cadden-Zimansky, Bolotin, and Kim}}]{Ghahari11}
\bibinfo{author}{\bibfnamefont{F.}~\bibnamefont{Ghahari}},
  \bibinfo{author}{\bibfnamefont{Y.}~\bibnamefont{Zhao}},
  \bibinfo{author}{\bibfnamefont{P.}~\bibnamefont{Cadden-Zimansky}},
  \bibinfo{author}{\bibfnamefont{K.}~\bibnamefont{Bolotin}}, \bibnamefont{and}
  \bibinfo{author}{\bibfnamefont{P.}~\bibnamefont{Kim}},
  \bibinfo{journal}{Phys. Rev. Lett.} \textbf{\bibinfo{volume}{106}},
  \bibinfo{pages}{046801} (\bibinfo{year}{2011}),
  \urlprefix\url{https://link.aps.org/doi/10.1103/PhysRevLett.106.046801}.

\bibitem[{\citenamefont{Dean et~al.}(2011)\citenamefont{Dean, Young,
  Cadden-Zimansky, Wang, Ren, Watanabe, Taniguchi, Kim, Hone, and
  Shepard}}]{Dean11}
\bibinfo{author}{\bibfnamefont{C.~R.} \bibnamefont{Dean}},
  \bibinfo{author}{\bibfnamefont{A.~F.} \bibnamefont{Young}},
  \bibinfo{author}{\bibfnamefont{P.}~\bibnamefont{Cadden-Zimansky}},
  \bibinfo{author}{\bibfnamefont{L.}~\bibnamefont{Wang}},
  \bibinfo{author}{\bibfnamefont{H.}~\bibnamefont{Ren}},
  \bibinfo{author}{\bibfnamefont{K.}~\bibnamefont{Watanabe}},
  \bibinfo{author}{\bibfnamefont{T.}~\bibnamefont{Taniguchi}},
  \bibinfo{author}{\bibfnamefont{P.}~\bibnamefont{Kim}},
  \bibinfo{author}{\bibfnamefont{J.}~\bibnamefont{Hone}}, \bibnamefont{and}
  \bibinfo{author}{\bibfnamefont{K.~L.} \bibnamefont{Shepard}},
  \bibinfo{journal}{Nature Physics} \textbf{\bibinfo{volume}{7}},
  \bibinfo{pages}{693} (\bibinfo{year}{2011}).

\bibitem[{\citenamefont{Feldman et~al.}(2012)\citenamefont{Feldman, Krauss,
  Smet, and Yacoby}}]{Feldman12}
\bibinfo{author}{\bibfnamefont{B.~E.} \bibnamefont{Feldman}},
  \bibinfo{author}{\bibfnamefont{B.}~\bibnamefont{Krauss}},
  \bibinfo{author}{\bibfnamefont{J.~H.} \bibnamefont{Smet}}, \bibnamefont{and}
  \bibinfo{author}{\bibfnamefont{A.}~\bibnamefont{Yacoby}},
  \bibinfo{journal}{Science} \textbf{\bibinfo{volume}{337}},
  \bibinfo{pages}{1196} (\bibinfo{year}{2012}),
  \eprint{http://www.sciencemag.org/content/337/6099/1196.full.pdf},
  \urlprefix\url{http://www.sciencemag.org/content/337/6099/1196.abstract}.

\bibitem[{\citenamefont{Feldman et~al.}(2013)\citenamefont{Feldman, Levin,
  Krauss, Abanin, Halperin, Smet, and Yacoby}}]{Feldman13}
\bibinfo{author}{\bibfnamefont{B.~E.} \bibnamefont{Feldman}},
  \bibinfo{author}{\bibfnamefont{A.~J.} \bibnamefont{Levin}},
  \bibinfo{author}{\bibfnamefont{B.}~\bibnamefont{Krauss}},
  \bibinfo{author}{\bibfnamefont{D.~A.} \bibnamefont{Abanin}},
  \bibinfo{author}{\bibfnamefont{B.~I.} \bibnamefont{Halperin}},
  \bibinfo{author}{\bibfnamefont{J.~H.} \bibnamefont{Smet}}, \bibnamefont{and}
  \bibinfo{author}{\bibfnamefont{A.}~\bibnamefont{Yacoby}},
  \bibinfo{journal}{Phys. Rev. Lett.} \textbf{\bibinfo{volume}{111}},
  \bibinfo{pages}{076802} (\bibinfo{year}{2013}),
  \urlprefix\url{http://link.aps.org/doi/10.1103/PhysRevLett.111.076802}.

\bibitem[{\citenamefont{Amet et~al.}(2015)\citenamefont{Amet, Bestwick,
  Williams, Balicas, Watanabe, Taniguchi, and Goldhaber-Gordon}}]{Amet15}
\bibinfo{author}{\bibfnamefont{F.}~\bibnamefont{Amet}},
  \bibinfo{author}{\bibfnamefont{A.~J.} \bibnamefont{Bestwick}},
  \bibinfo{author}{\bibfnamefont{J.~R.} \bibnamefont{Williams}},
  \bibinfo{author}{\bibfnamefont{L.}~\bibnamefont{Balicas}},
  \bibinfo{author}{\bibfnamefont{K.}~\bibnamefont{Watanabe}},
  \bibinfo{author}{\bibfnamefont{T.}~\bibnamefont{Taniguchi}},
  \bibnamefont{and}
  \bibinfo{author}{\bibfnamefont{D.}~\bibnamefont{Goldhaber-Gordon}},
  \bibinfo{journal}{Nat. Commun.} \textbf{\bibinfo{volume}{6}},
  \bibinfo{pages}{5838} (\bibinfo{year}{2015}),
  \urlprefix\url{http://dx.doi.org/10.1038/ncomms6838}.

\bibitem[{\citenamefont{Kim et~al.}(2019)\citenamefont{Kim, Balram, Taniguchi,
  Watanabe, Jain, and Smet}}]{Kim19}
\bibinfo{author}{\bibfnamefont{Y.}~\bibnamefont{Kim}},
  \bibinfo{author}{\bibfnamefont{A.~C.} \bibnamefont{Balram}},
  \bibinfo{author}{\bibfnamefont{T.}~\bibnamefont{Taniguchi}},
  \bibinfo{author}{\bibfnamefont{K.}~\bibnamefont{Watanabe}},
  \bibinfo{author}{\bibfnamefont{J.~K.} \bibnamefont{Jain}}, \bibnamefont{and}
  \bibinfo{author}{\bibfnamefont{J.~H.} \bibnamefont{Smet}},
  \bibinfo{journal}{Nature Physics} \textbf{\bibinfo{volume}{15}},
  \bibinfo{pages}{154} (\bibinfo{year}{2019}), ISSN \bibinfo{issn}{1745-2481},
  \urlprefix\url{https://doi.org/10.1038/s41567-018-0355-x}.

\bibitem[{\citenamefont{Balram et~al.}(2015{\natexlab{a}})\citenamefont{Balram,
  T\"oke, W\'ojs, and Jain}}]{Balram15}
\bibinfo{author}{\bibfnamefont{A.~C.} \bibnamefont{Balram}},
  \bibinfo{author}{\bibfnamefont{C.}~\bibnamefont{T\"oke}},
  \bibinfo{author}{\bibfnamefont{A.}~\bibnamefont{W\'ojs}}, \bibnamefont{and}
  \bibinfo{author}{\bibfnamefont{J.~K.} \bibnamefont{Jain}},
  \bibinfo{journal}{Phys. Rev. B} \textbf{\bibinfo{volume}{91}},
  \bibinfo{pages}{045109} (\bibinfo{year}{2015}{\natexlab{a}}),
  \urlprefix\url{http://link.aps.org/doi/10.1103/PhysRevB.91.045109}.

\bibitem[{\citenamefont{Balram}(2016)}]{Balram16c}
\bibinfo{author}{\bibfnamefont{A.~C.} \bibnamefont{Balram}},
  \bibinfo{journal}{Phys. Rev. B} \textbf{\bibinfo{volume}{94}},
  \bibinfo{pages}{165303} (\bibinfo{year}{2016}),
  \urlprefix\url{http://link.aps.org/doi/10.1103/PhysRevB.94.165303}.

\bibitem[{\citenamefont{Mandal and Jain}(2002)}]{Mandal02}
\bibinfo{author}{\bibfnamefont{S.~S.} \bibnamefont{Mandal}} \bibnamefont{and}
  \bibinfo{author}{\bibfnamefont{J.~K.} \bibnamefont{Jain}},
  \bibinfo{journal}{Phys. Rev. B} \textbf{\bibinfo{volume}{66}},
  \bibinfo{pages}{155302} (\bibinfo{year}{2002}),
  \urlprefix\url{http://link.aps.org/doi/10.1103/PhysRevB.66.155302}.

\bibitem[{\citenamefont{Mukherjee et~al.}(2014)\citenamefont{Mukherjee, Mandal,
  Wu, W\'ojs, and Jain}}]{Mukherjee14}
\bibinfo{author}{\bibfnamefont{S.}~\bibnamefont{Mukherjee}},
  \bibinfo{author}{\bibfnamefont{S.~S.} \bibnamefont{Mandal}},
  \bibinfo{author}{\bibfnamefont{Y.-H.} \bibnamefont{Wu}},
  \bibinfo{author}{\bibfnamefont{A.}~\bibnamefont{W\'ojs}}, \bibnamefont{and}
  \bibinfo{author}{\bibfnamefont{J.~K.} \bibnamefont{Jain}},
  \bibinfo{journal}{Phys. Rev. Lett.} \textbf{\bibinfo{volume}{112}},
  \bibinfo{pages}{016801} (\bibinfo{year}{2014}),
  \urlprefix\url{http://link.aps.org/doi/10.1103/PhysRevLett.112.016801}.

\bibitem[{\citenamefont{Sitko et~al.}(1996)\citenamefont{Sitko, Yi, Yi, and
  Quinn}}]{Sitko96}
\bibinfo{author}{\bibfnamefont{P.}~\bibnamefont{Sitko}},
  \bibinfo{author}{\bibfnamefont{S.~N.} \bibnamefont{Yi}},
  \bibinfo{author}{\bibfnamefont{K.~S.} \bibnamefont{Yi}}, \bibnamefont{and}
  \bibinfo{author}{\bibfnamefont{J.~J.} \bibnamefont{Quinn}},
  \bibinfo{journal}{Phys. Rev. Lett.} \textbf{\bibinfo{volume}{76}},
  \bibinfo{pages}{3396} (\bibinfo{year}{1996}),
  \urlprefix\url{http://link.aps.org/doi/10.1103/PhysRevLett.76.3396}.

\bibitem[{\citenamefont{W\'ojs and Quinn}(2000)}]{Wojs00}
\bibinfo{author}{\bibfnamefont{A.}~\bibnamefont{W\'ojs}} \bibnamefont{and}
  \bibinfo{author}{\bibfnamefont{J.~J.} \bibnamefont{Quinn}},
  \bibinfo{journal}{Phys. Rev. B} \textbf{\bibinfo{volume}{61}},
  \bibinfo{pages}{2846} (\bibinfo{year}{2000}),
  \urlprefix\url{http://link.aps.org/doi/10.1103/PhysRevB.61.2846}.

\bibitem[{\citenamefont{W\'ojs et~al.}(2004)\citenamefont{W\'ojs, Yi, and
  Quinn}}]{Wojs04}
\bibinfo{author}{\bibfnamefont{A.}~\bibnamefont{W\'ojs}},
  \bibinfo{author}{\bibfnamefont{K.-S.} \bibnamefont{Yi}}, \bibnamefont{and}
  \bibinfo{author}{\bibfnamefont{J.~J.} \bibnamefont{Quinn}},
  \bibinfo{journal}{Phys. Rev. B} \textbf{\bibinfo{volume}{69}},
  \bibinfo{pages}{205322} (\bibinfo{year}{2004}),
  \urlprefix\url{http://link.aps.org/doi/10.1103/PhysRevB.69.205322}.

\bibitem[{\citenamefont{Lee et~al.}(2001)\citenamefont{Lee, Scarola, and
  Jain}}]{Lee01}
\bibinfo{author}{\bibfnamefont{S.-Y.} \bibnamefont{Lee}},
  \bibinfo{author}{\bibfnamefont{V.~W.} \bibnamefont{Scarola}},
  \bibnamefont{and} \bibinfo{author}{\bibfnamefont{J.~K.} \bibnamefont{Jain}},
  \bibinfo{journal}{Phys. Rev. Lett.} \textbf{\bibinfo{volume}{87}},
  \bibinfo{pages}{256803} (\bibinfo{year}{2001}),
  \urlprefix\url{http://link.aps.org/doi/10.1103/PhysRevLett.87.256803}.

\bibitem[{\citenamefont{Lee et~al.}(2002)\citenamefont{Lee, Scarola, and
  Jain}}]{Lee02}
\bibinfo{author}{\bibfnamefont{S.-Y.} \bibnamefont{Lee}},
  \bibinfo{author}{\bibfnamefont{V.~W.} \bibnamefont{Scarola}},
  \bibnamefont{and} \bibinfo{author}{\bibfnamefont{J.~K.} \bibnamefont{Jain}},
  \bibinfo{journal}{Phys. Rev. B} \textbf{\bibinfo{volume}{66}},
  \bibinfo{pages}{085336} (\bibinfo{year}{2002}),
  \urlprefix\url{http://link.aps.org/doi/10.1103/PhysRevB.66.085336}.

\bibitem[{\citenamefont{Haldane}(1983)}]{Haldane83}
\bibinfo{author}{\bibfnamefont{F.~D.~M.} \bibnamefont{Haldane}},
  \bibinfo{journal}{Phys. Rev. Lett.} \textbf{\bibinfo{volume}{51}},
  \bibinfo{pages}{605} (\bibinfo{year}{1983}),
  \urlprefix\url{http://link.aps.org/doi/10.1103/PhysRevLett.51.605}.

\bibitem[{\citenamefont{Wen and Zee}(1992)}]{Wen92}
\bibinfo{author}{\bibfnamefont{X.~G.} \bibnamefont{Wen}} \bibnamefont{and}
  \bibinfo{author}{\bibfnamefont{A.}~\bibnamefont{Zee}},
  \bibinfo{journal}{Phys. Rev. Lett.} \textbf{\bibinfo{volume}{69}},
  \bibinfo{pages}{953} (\bibinfo{year}{1992}),
  \urlprefix\url{http://link.aps.org/doi/10.1103/PhysRevLett.69.953}.

\bibitem[{\citenamefont{Mukherjee and
  Mandal}(2015{\natexlab{a}})}]{Mukherjee15b}
\bibinfo{author}{\bibfnamefont{S.}~\bibnamefont{Mukherjee}} \bibnamefont{and}
  \bibinfo{author}{\bibfnamefont{S.~S.} \bibnamefont{Mandal}},
  \bibinfo{journal}{Phys. Rev. B} \textbf{\bibinfo{volume}{92}},
  \bibinfo{pages}{235302} (\bibinfo{year}{2015}{\natexlab{a}}),
  \urlprefix\url{http://link.aps.org/doi/10.1103/PhysRevB.92.235302}.

\bibitem[{\citenamefont{Das et~al.}(2021)\citenamefont{Das, Das, and
  Mandal}}]{Das20}
\bibinfo{author}{\bibfnamefont{S.}~\bibnamefont{Das}},
  \bibinfo{author}{\bibfnamefont{S.}~\bibnamefont{Das}}, \bibnamefont{and}
  \bibinfo{author}{\bibfnamefont{S.~S.} \bibnamefont{Mandal}},
  \bibinfo{journal}{Phys. Rev. B} \textbf{\bibinfo{volume}{103}},
  \bibinfo{pages}{075304} (\bibinfo{year}{2021}),
  \urlprefix\url{https://link.aps.org/doi/10.1103/PhysRevB.103.075304}.

\bibitem[{\citenamefont{Regnault et~al.}(2017)\citenamefont{Regnault, Maciejko,
  Kivelson, and Sondhi}}]{Regnault16}
\bibinfo{author}{\bibfnamefont{N.}~\bibnamefont{Regnault}},
  \bibinfo{author}{\bibfnamefont{J.}~\bibnamefont{Maciejko}},
  \bibinfo{author}{\bibfnamefont{S.~A.} \bibnamefont{Kivelson}},
  \bibnamefont{and} \bibinfo{author}{\bibfnamefont{S.~L.}
  \bibnamefont{Sondhi}}, \bibinfo{journal}{Phys. Rev. B}
  \textbf{\bibinfo{volume}{96}}, \bibinfo{pages}{035150}
  (\bibinfo{year}{2017}),
  \urlprefix\url{https://link.aps.org/doi/10.1103/PhysRevB.96.035150}.

\bibitem[{\citenamefont{Misguich et~al.}(2020)\citenamefont{Misguich,
  Jolicoeur, and Mizusaki}}]{Jolicoeur17a}
\bibinfo{author}{\bibfnamefont{G.}~\bibnamefont{Misguich}},
  \bibinfo{author}{\bibfnamefont{T.}~\bibnamefont{Jolicoeur}},
  \bibnamefont{and} \bibinfo{author}{\bibfnamefont{T.}~\bibnamefont{Mizusaki}},
  \bibinfo{journal}{Phys. Rev. B} \textbf{\bibinfo{volume}{102}},
  \bibinfo{pages}{245107} (\bibinfo{year}{2020}),
  \urlprefix\url{https://link.aps.org/doi/10.1103/PhysRevB.102.245107}.

\bibitem[{\citenamefont{Balram and Jain}(2017{\natexlab{a}})}]{Balram17b}
\bibinfo{author}{\bibfnamefont{A.~C.} \bibnamefont{Balram}} \bibnamefont{and}
  \bibinfo{author}{\bibfnamefont{J.~K.} \bibnamefont{Jain}},
  \bibinfo{journal}{Phys. Rev. B} \textbf{\bibinfo{volume}{96}},
  \bibinfo{pages}{245142} (\bibinfo{year}{2017}{\natexlab{a}}),
  \urlprefix\url{http://link.aps.org/doi/10.1103/PhysRevB.96.245142}.

\bibitem[{\citenamefont{Jain}(1989{\natexlab{b}})}]{Jain89b}
\bibinfo{author}{\bibfnamefont{J.~K.} \bibnamefont{Jain}},
  \bibinfo{journal}{Phys. Rev. B} \textbf{\bibinfo{volume}{40}},
  \bibinfo{pages}{8079} (\bibinfo{year}{1989}{\natexlab{b}}),
  \urlprefix\url{http://link.aps.org/doi/10.1103/PhysRevB.40.8079}.

\bibitem[{\citenamefont{Balram}(2021{\natexlab{a}})}]{Balram21c}
\bibinfo{author}{\bibfnamefont{A.~C.} \bibnamefont{Balram}},
  \bibinfo{journal}{Phys. Rev. B} \textbf{\bibinfo{volume}{103}},
  \bibinfo{pages}{155103} (\bibinfo{year}{2021}{\natexlab{a}}),
  \urlprefix\url{https://link.aps.org/doi/10.1103/PhysRevB.103.155103}.

\bibitem[{\citenamefont{Wen}(1991{\natexlab{a}})}]{Wen91}
\bibinfo{author}{\bibfnamefont{X.~G.} \bibnamefont{Wen}},
  \bibinfo{journal}{Phys. Rev. Lett.} \textbf{\bibinfo{volume}{66}},
  \bibinfo{pages}{802} (\bibinfo{year}{1991}{\natexlab{a}}),
  \urlprefix\url{http://link.aps.org/doi/10.1103/PhysRevLett.66.802}.

\bibitem[{\citenamefont{Laughlin}(1983)}]{Laughlin83}
\bibinfo{author}{\bibfnamefont{R.~B.} \bibnamefont{Laughlin}},
  \bibinfo{journal}{Phys. Rev. Lett.} \textbf{\bibinfo{volume}{50}},
  \bibinfo{pages}{1395} (\bibinfo{year}{1983}),
  \urlprefix\url{http://link.aps.org/doi/10.1103/PhysRevLett.50.1395}.

\bibitem[{\citenamefont{Wu et~al.}(2017)\citenamefont{Wu, Shi, and
  Jain}}]{Wu17}
\bibinfo{author}{\bibfnamefont{Y.}~\bibnamefont{Wu}},
  \bibinfo{author}{\bibfnamefont{T.}~\bibnamefont{Shi}}, \bibnamefont{and}
  \bibinfo{author}{\bibfnamefont{J.~K.} \bibnamefont{Jain}},
  \bibinfo{journal}{Nano Letters} \textbf{\bibinfo{volume}{17}},
  \bibinfo{pages}{4643} (\bibinfo{year}{2017}), \bibinfo{note}{pMID: 28649831},
  \eprint{http://dx.doi.org/10.1021/acs.nanolett.7b01080},
  \urlprefix\url{http://dx.doi.org/10.1021/acs.nanolett.7b01080}.

\bibitem[{\citenamefont{Balram et~al.}(2018{\natexlab{a}})\citenamefont{Balram,
  Barkeshli, and Rudner}}]{Balram18}
\bibinfo{author}{\bibfnamefont{A.~C.} \bibnamefont{Balram}},
  \bibinfo{author}{\bibfnamefont{M.}~\bibnamefont{Barkeshli}},
  \bibnamefont{and} \bibinfo{author}{\bibfnamefont{M.~S.}
  \bibnamefont{Rudner}}, \bibinfo{journal}{Phys. Rev. B}
  \textbf{\bibinfo{volume}{98}}, \bibinfo{pages}{035127}
  (\bibinfo{year}{2018}{\natexlab{a}}),
  \urlprefix\url{https://link.aps.org/doi/10.1103/PhysRevB.98.035127}.

\bibitem[{\citenamefont{Balram et~al.}(2018{\natexlab{b}})\citenamefont{Balram,
  Mukherjee, Park, Barkeshli, Rudner, and Jain}}]{Balram18a}
\bibinfo{author}{\bibfnamefont{A.~C.} \bibnamefont{Balram}},
  \bibinfo{author}{\bibfnamefont{S.}~\bibnamefont{Mukherjee}},
  \bibinfo{author}{\bibfnamefont{K.}~\bibnamefont{Park}},
  \bibinfo{author}{\bibfnamefont{M.}~\bibnamefont{Barkeshli}},
  \bibinfo{author}{\bibfnamefont{M.~S.} \bibnamefont{Rudner}},
  \bibnamefont{and} \bibinfo{author}{\bibfnamefont{J.~K.} \bibnamefont{Jain}},
  \bibinfo{journal}{Phys. Rev. Lett.} \textbf{\bibinfo{volume}{121}},
  \bibinfo{pages}{186601} (\bibinfo{year}{2018}{\natexlab{b}}),
  \urlprefix\url{https://link.aps.org/doi/10.1103/PhysRevLett.121.186601}.

\bibitem[{\citenamefont{Balram et~al.}(2019)\citenamefont{Balram, Barkeshli,
  and Rudner}}]{Balram19}
\bibinfo{author}{\bibfnamefont{A.~C.} \bibnamefont{Balram}},
  \bibinfo{author}{\bibfnamefont{M.}~\bibnamefont{Barkeshli}},
  \bibnamefont{and} \bibinfo{author}{\bibfnamefont{M.~S.}
  \bibnamefont{Rudner}}, \bibinfo{journal}{Phys. Rev. B}
  \textbf{\bibinfo{volume}{99}}, \bibinfo{pages}{241108}
  (\bibinfo{year}{2019}),
  \urlprefix\url{https://link.aps.org/doi/10.1103/PhysRevB.99.241108}.

\bibitem[{\citenamefont{Bandyopadhyay et~al.}(2018)\citenamefont{Bandyopadhyay,
  Chen, Ahari, Ortiz, Nussinov, and Seidel}}]{Bandyopadhyay18}
\bibinfo{author}{\bibfnamefont{S.}~\bibnamefont{Bandyopadhyay}},
  \bibinfo{author}{\bibfnamefont{L.}~\bibnamefont{Chen}},
  \bibinfo{author}{\bibfnamefont{M.~T.} \bibnamefont{Ahari}},
  \bibinfo{author}{\bibfnamefont{G.}~\bibnamefont{Ortiz}},
  \bibinfo{author}{\bibfnamefont{Z.}~\bibnamefont{Nussinov}}, \bibnamefont{and}
  \bibinfo{author}{\bibfnamefont{A.}~\bibnamefont{Seidel}},
  \bibinfo{journal}{Phys. Rev. B} \textbf{\bibinfo{volume}{98}},
  \bibinfo{pages}{161118} (\bibinfo{year}{2018}),
  \urlprefix\url{https://link.aps.org/doi/10.1103/PhysRevB.98.161118}.

\bibitem[{\citenamefont{Faugno et~al.}(2019)\citenamefont{Faugno, Balram,
  Barkeshli, and Jain}}]{Faugno19}
\bibinfo{author}{\bibfnamefont{W.~N.} \bibnamefont{Faugno}},
  \bibinfo{author}{\bibfnamefont{A.~C.} \bibnamefont{Balram}},
  \bibinfo{author}{\bibfnamefont{M.}~\bibnamefont{Barkeshli}},
  \bibnamefont{and} \bibinfo{author}{\bibfnamefont{J.~K.} \bibnamefont{Jain}},
  \bibinfo{journal}{Phys. Rev. Lett.} \textbf{\bibinfo{volume}{123}},
  \bibinfo{pages}{016802} (\bibinfo{year}{2019}),
  \urlprefix\url{https://link.aps.org/doi/10.1103/PhysRevLett.123.016802}.

\bibitem[{\citenamefont{Balram et~al.}(2020)\citenamefont{Balram, Jain, and
  Barkeshli}}]{Balram20}
\bibinfo{author}{\bibfnamefont{A.~C.} \bibnamefont{Balram}},
  \bibinfo{author}{\bibfnamefont{J.~K.} \bibnamefont{Jain}}, \bibnamefont{and}
  \bibinfo{author}{\bibfnamefont{M.}~\bibnamefont{Barkeshli}},
  \bibinfo{journal}{Phys. Rev. Research} \textbf{\bibinfo{volume}{2}},
  \bibinfo{pages}{013349} (\bibinfo{year}{2020}),
  \urlprefix\url{https://link.aps.org/doi/10.1103/PhysRevResearch.2.013349}.

\bibitem[{\citenamefont{Balram}(2021{\natexlab{b}})}]{Balram20a}
\bibinfo{author}{\bibfnamefont{A.~C.} \bibnamefont{Balram}},
  \bibinfo{journal}{SciPost Phys.} \textbf{\bibinfo{volume}{10}},
  \bibinfo{pages}{83} (\bibinfo{year}{2021}{\natexlab{b}}),
  \urlprefix\url{https://scipost.org/10.21468/SciPostPhys.10.4.083}.

\bibitem[{\citenamefont{Balram and W\'ojs}(2020)}]{Balram20b}
\bibinfo{author}{\bibfnamefont{A.~C.} \bibnamefont{Balram}} \bibnamefont{and}
  \bibinfo{author}{\bibfnamefont{A.}~\bibnamefont{W\'ojs}},
  \bibinfo{journal}{Phys. Rev. Research} \textbf{\bibinfo{volume}{2}},
  \bibinfo{pages}{032035} (\bibinfo{year}{2020}),
  \urlprefix\url{https://link.aps.org/doi/10.1103/PhysRevResearch.2.032035}.

\bibitem[{\citenamefont{Faugno et~al.}(2020)\citenamefont{Faugno, Jain, and
  Balram}}]{Faugno20a}
\bibinfo{author}{\bibfnamefont{W.~N.} \bibnamefont{Faugno}},
  \bibinfo{author}{\bibfnamefont{J.~K.} \bibnamefont{Jain}}, \bibnamefont{and}
  \bibinfo{author}{\bibfnamefont{A.~C.} \bibnamefont{Balram}},
  \bibinfo{journal}{Phys. Rev. Research} \textbf{\bibinfo{volume}{2}},
  \bibinfo{pages}{033223} (\bibinfo{year}{2020}),
  \urlprefix\url{https://link.aps.org/doi/10.1103/PhysRevResearch.2.033223}.

\bibitem[{\citenamefont{Faugno et~al.}(2021)\citenamefont{Faugno, Zhao, Balram,
  Jolicoeur, and Jain}}]{Faugno21}
\bibinfo{author}{\bibfnamefont{W.~N.} \bibnamefont{Faugno}},
  \bibinfo{author}{\bibfnamefont{T.}~\bibnamefont{Zhao}},
  \bibinfo{author}{\bibfnamefont{A.~C.} \bibnamefont{Balram}},
  \bibinfo{author}{\bibfnamefont{T.}~\bibnamefont{Jolicoeur}},
  \bibnamefont{and} \bibinfo{author}{\bibfnamefont{J.~K.} \bibnamefont{Jain}},
  \bibinfo{journal}{Phys. Rev. B} \textbf{\bibinfo{volume}{103}},
  \bibinfo{pages}{085303} (\bibinfo{year}{2021}),
  \urlprefix\url{https://link.aps.org/doi/10.1103/PhysRevB.103.085303}.

\bibitem[{\citenamefont{Balram}(2021{\natexlab{c}})}]{Balram21}
\bibinfo{author}{\bibfnamefont{A.~C.} \bibnamefont{Balram}},
  \bibinfo{journal}{SciPost Phys.} \textbf{\bibinfo{volume}{10}},
  \bibinfo{pages}{83} (\bibinfo{year}{2021}{\natexlab{c}}),
  \urlprefix\url{https://scipost.org/10.21468/SciPostPhys.10.4.083}.

\bibitem[{\citenamefont{Balram and W\'ojs}(2021)}]{Balram21a}
\bibinfo{author}{\bibfnamefont{A.~C.} \bibnamefont{Balram}} \bibnamefont{and}
  \bibinfo{author}{\bibfnamefont{A.}~\bibnamefont{W\'ojs}},
  \bibinfo{journal}{Phys. Rev. Research} \textbf{\bibinfo{volume}{3}},
  \bibinfo{pages}{033087} (\bibinfo{year}{2021}),
  \urlprefix\url{https://link.aps.org/doi/10.1103/PhysRevResearch.3.033087}.

\bibitem[{\citenamefont{Balram}(2022)}]{Balram21b}
\bibinfo{author}{\bibfnamefont{A.~C.} \bibnamefont{Balram}},
  \bibinfo{journal}{Phys. Rev. B} \textbf{\bibinfo{volume}{105}},
  \bibinfo{pages}{L121406} (\bibinfo{year}{2022}),
  \urlprefix\url{https://link.aps.org/doi/10.1103/PhysRevB.105.L121406}.

\bibitem[{\citenamefont{Balram et~al.}(2015{\natexlab{b}})\citenamefont{Balram,
  T\"oke, W\'ojs, and Jain}}]{Balram15a}
\bibinfo{author}{\bibfnamefont{A.~C.} \bibnamefont{Balram}},
  \bibinfo{author}{\bibfnamefont{C.}~\bibnamefont{T\"oke}},
  \bibinfo{author}{\bibfnamefont{A.}~\bibnamefont{W\'ojs}}, \bibnamefont{and}
  \bibinfo{author}{\bibfnamefont{J.~K.} \bibnamefont{Jain}},
  \bibinfo{journal}{Phys. Rev. B} \textbf{\bibinfo{volume}{92}},
  \bibinfo{pages}{075410} (\bibinfo{year}{2015}{\natexlab{b}}),
  \urlprefix\url{http://link.aps.org/doi/10.1103/PhysRevB.92.075410}.

\bibitem[{\citenamefont{Balram and Jain}(2016)}]{Balram16b}
\bibinfo{author}{\bibfnamefont{A.~C.} \bibnamefont{Balram}} \bibnamefont{and}
  \bibinfo{author}{\bibfnamefont{J.~K.} \bibnamefont{Jain}},
  \bibinfo{journal}{Phys. Rev. B} \textbf{\bibinfo{volume}{93}},
  \bibinfo{pages}{235152} (\bibinfo{year}{2016}),
  \urlprefix\url{http://link.aps.org/doi/10.1103/PhysRevB.93.235152}.

\bibitem[{\citenamefont{Jain and Kamilla}(1997)}]{Jain97}
\bibinfo{author}{\bibfnamefont{J.~K.} \bibnamefont{Jain}} \bibnamefont{and}
  \bibinfo{author}{\bibfnamefont{R.~K.} \bibnamefont{Kamilla}},
  \bibinfo{journal}{Int. J. Mod. Phys. B} \textbf{\bibinfo{volume}{11}},
  \bibinfo{pages}{2621} (\bibinfo{year}{1997}).

\bibitem[{\citenamefont{Davenport and Simon}(2012)}]{Davenport12}
\bibinfo{author}{\bibfnamefont{S.~C.} \bibnamefont{Davenport}}
  \bibnamefont{and} \bibinfo{author}{\bibfnamefont{S.~H.} \bibnamefont{Simon}},
  \bibinfo{journal}{Phys. Rev. B} \textbf{\bibinfo{volume}{85}},
  \bibinfo{pages}{245303} (\bibinfo{year}{2012}),
  \urlprefix\url{http://link.aps.org/doi/10.1103/PhysRevB.85.245303}.

\bibitem[{\citenamefont{Morf et~al.}(1986)\citenamefont{Morf, d'Ambrumenil, and
  Halperin}}]{Morf86}
\bibinfo{author}{\bibfnamefont{R.}~\bibnamefont{Morf}},
  \bibinfo{author}{\bibfnamefont{N.}~\bibnamefont{d'Ambrumenil}},
  \bibnamefont{and} \bibinfo{author}{\bibfnamefont{B.~I.}
  \bibnamefont{Halperin}}, \bibinfo{journal}{Phys. Rev. B}
  \textbf{\bibinfo{volume}{34}}, \bibinfo{pages}{3037} (\bibinfo{year}{1986}),
  \urlprefix\url{http://link.aps.org/doi/10.1103/PhysRevB.34.3037}.

\bibitem[{\citenamefont{Balram et~al.}(2015{\natexlab{c}})\citenamefont{Balram,
  T\ifmmode~\mbox{\H{o}}\else \H{o}\fi{}ke, and Jain}}]{Balram15b}
\bibinfo{author}{\bibfnamefont{A.~C.} \bibnamefont{Balram}},
  \bibinfo{author}{\bibfnamefont{C.}~\bibnamefont{T\ifmmode~\mbox{\H{o}}\else
  \H{o}\fi{}ke}}, \bibnamefont{and} \bibinfo{author}{\bibfnamefont{J.~K.}
  \bibnamefont{Jain}}, \bibinfo{journal}{Phys. Rev. Lett.}
  \textbf{\bibinfo{volume}{115}}, \bibinfo{pages}{186805}
  (\bibinfo{year}{2015}{\natexlab{c}}),
  \urlprefix\url{http://link.aps.org/doi/10.1103/PhysRevLett.115.186805}.

\bibitem[{\citenamefont{Balram and Jain}(2017{\natexlab{b}})}]{Balram17}
\bibinfo{author}{\bibfnamefont{A.~C.} \bibnamefont{Balram}} \bibnamefont{and}
  \bibinfo{author}{\bibfnamefont{J.~K.} \bibnamefont{Jain}},
  \bibinfo{journal}{Phys. Rev. B} \textbf{\bibinfo{volume}{96}},
  \bibinfo{pages}{235102} (\bibinfo{year}{2017}{\natexlab{b}}),
  \urlprefix\url{http://link.aps.org/doi/10.1103/PhysRevB.96.235102}.

\bibitem[{\citenamefont{Read and Rezayi}(1999)}]{Read99}
\bibinfo{author}{\bibfnamefont{N.}~\bibnamefont{Read}} \bibnamefont{and}
  \bibinfo{author}{\bibfnamefont{E.}~\bibnamefont{Rezayi}},
  \bibinfo{journal}{Phys. Rev. B} \textbf{\bibinfo{volume}{59}},
  \bibinfo{pages}{8084} (\bibinfo{year}{1999}),
  \urlprefix\url{http://link.aps.org/doi/10.1103/PhysRevB.59.8084}.

\bibitem[{\citenamefont{Hutasoit et~al.}(2017)\citenamefont{Hutasoit, Balram,
  Mukherjee, Wu, Mandal, W\'ojs, Cheianov, and Jain}}]{Hutasoit16}
\bibinfo{author}{\bibfnamefont{J.~A.} \bibnamefont{Hutasoit}},
  \bibinfo{author}{\bibfnamefont{A.~C.} \bibnamefont{Balram}},
  \bibinfo{author}{\bibfnamefont{S.}~\bibnamefont{Mukherjee}},
  \bibinfo{author}{\bibfnamefont{Y.-H.} \bibnamefont{Wu}},
  \bibinfo{author}{\bibfnamefont{S.~S.} \bibnamefont{Mandal}},
  \bibinfo{author}{\bibfnamefont{A.}~\bibnamefont{W\'ojs}},
  \bibinfo{author}{\bibfnamefont{V.}~\bibnamefont{Cheianov}}, \bibnamefont{and}
  \bibinfo{author}{\bibfnamefont{J.~K.} \bibnamefont{Jain}},
  \bibinfo{journal}{Phys. Rev. B} \textbf{\bibinfo{volume}{95}},
  \bibinfo{pages}{125302} (\bibinfo{year}{2017}),
  \urlprefix\url{http://link.aps.org/doi/10.1103/PhysRevB.95.125302}.

\bibitem[{\citenamefont{Kamilla et~al.}(1996)\citenamefont{Kamilla, Wu, and
  Jain}}]{Kamilla96b}
\bibinfo{author}{\bibfnamefont{R.~K.} \bibnamefont{Kamilla}},
  \bibinfo{author}{\bibfnamefont{X.~G.} \bibnamefont{Wu}}, \bibnamefont{and}
  \bibinfo{author}{\bibfnamefont{J.~K.} \bibnamefont{Jain}},
  \bibinfo{journal}{Phys. Rev. B} \textbf{\bibinfo{volume}{54}},
  \bibinfo{pages}{4873} (\bibinfo{year}{1996}),
  \urlprefix\url{http://link.aps.org/doi/10.1103/PhysRevB.54.4873}.

\bibitem[{\citenamefont{Balram and Pu}(2017)}]{Balram16d}
\bibinfo{author}{\bibfnamefont{A.~C.} \bibnamefont{Balram}} \bibnamefont{and}
  \bibinfo{author}{\bibfnamefont{S.}~\bibnamefont{Pu}}, \bibinfo{journal}{The
  European Physical Journal B} \textbf{\bibinfo{volume}{90}},
  \bibinfo{pages}{124} (\bibinfo{year}{2017}), ISSN \bibinfo{issn}{1434-6036},
  \urlprefix\url{http://dx.doi.org/10.1140/epjb/e2017-80177-5}.

\bibitem[{\citenamefont{Mukherjee and
  Mandal}(2015{\natexlab{b}})}]{Mukherjee15}
\bibinfo{author}{\bibfnamefont{S.}~\bibnamefont{Mukherjee}} \bibnamefont{and}
  \bibinfo{author}{\bibfnamefont{S.~S.} \bibnamefont{Mandal}},
  \bibinfo{journal}{Phys. Rev. Lett.} \textbf{\bibinfo{volume}{114}},
  \bibinfo{pages}{156802} (\bibinfo{year}{2015}{\natexlab{b}}),
  \urlprefix\url{http://link.aps.org/doi/10.1103/PhysRevLett.114.156802}.

\bibitem[{\citenamefont{Girvin et~al.}(1985)\citenamefont{Girvin, MacDonald,
  and Platzman}}]{Girvin85}
\bibinfo{author}{\bibfnamefont{S.~M.} \bibnamefont{Girvin}},
  \bibinfo{author}{\bibfnamefont{A.~H.} \bibnamefont{MacDonald}},
  \bibnamefont{and} \bibinfo{author}{\bibfnamefont{P.~M.}
  \bibnamefont{Platzman}}, \bibinfo{journal}{Phys. Rev. Lett.}
  \textbf{\bibinfo{volume}{54}}, \bibinfo{pages}{581} (\bibinfo{year}{1985}),
  \urlprefix\url{http://link.aps.org/doi/10.1103/PhysRevLett.54.581}.

\bibitem[{\citenamefont{Girvin et~al.}(1986)\citenamefont{Girvin, MacDonald,
  and Platzman}}]{Girvin86}
\bibinfo{author}{\bibfnamefont{S.~M.} \bibnamefont{Girvin}},
  \bibinfo{author}{\bibfnamefont{A.~H.} \bibnamefont{MacDonald}},
  \bibnamefont{and} \bibinfo{author}{\bibfnamefont{P.~M.}
  \bibnamefont{Platzman}}, \bibinfo{journal}{Phys. Rev. B}
  \textbf{\bibinfo{volume}{33}}, \bibinfo{pages}{2481} (\bibinfo{year}{1986}),
  \urlprefix\url{http://link.aps.org/doi/10.1103/PhysRevB.33.2481}.

\bibitem[{\citenamefont{Balram et~al.}(2022)\citenamefont{Balram, Liu, Gromov,
  and Papi\ifmmode~\acute{c}\else \'{c}\fi{}}}]{Balram21d}
\bibinfo{author}{\bibfnamefont{A.~C.} \bibnamefont{Balram}},
  \bibinfo{author}{\bibfnamefont{Z.}~\bibnamefont{Liu}},
  \bibinfo{author}{\bibfnamefont{A.}~\bibnamefont{Gromov}}, \bibnamefont{and}
  \bibinfo{author}{\bibfnamefont{Z.}~\bibnamefont{Papi\ifmmode~\acute{c}\else
  \'{c}\fi{}}}, \bibinfo{journal}{Phys. Rev. X} \textbf{\bibinfo{volume}{12}},
  \bibinfo{pages}{021008} (\bibinfo{year}{2022}),
  \urlprefix\url{https://link.aps.org/doi/10.1103/PhysRevX.12.021008}.

\bibitem[{\citenamefont{Liou et~al.}(2019)\citenamefont{Liou, Haldane, Yang,
  and Rezayi}}]{Liou19}
\bibinfo{author}{\bibfnamefont{S.-F.} \bibnamefont{Liou}},
  \bibinfo{author}{\bibfnamefont{F.~D.~M.} \bibnamefont{Haldane}},
  \bibinfo{author}{\bibfnamefont{K.}~\bibnamefont{Yang}}, \bibnamefont{and}
  \bibinfo{author}{\bibfnamefont{E.~H.} \bibnamefont{Rezayi}},
  \bibinfo{journal}{Phys. Rev. Lett.} \textbf{\bibinfo{volume}{123}},
  \bibinfo{pages}{146801} (\bibinfo{year}{2019}),
  \urlprefix\url{https://link.aps.org/doi/10.1103/PhysRevLett.123.146801}.

\bibitem[{\citenamefont{L\'opez and Fradkin}(2004)}]{Lopez04}
\bibinfo{author}{\bibfnamefont{A.}~\bibnamefont{L\'opez}} \bibnamefont{and}
  \bibinfo{author}{\bibfnamefont{E.}~\bibnamefont{Fradkin}},
  \bibinfo{journal}{Phys. Rev. B} \textbf{\bibinfo{volume}{69}},
  \bibinfo{pages}{155322} (\bibinfo{year}{2004}),
  \urlprefix\url{https://link.aps.org/doi/10.1103/PhysRevB.69.155322}.

\bibitem[{\citenamefont{Wang and Yang}(2022)}]{Wang22}
\bibinfo{author}{\bibfnamefont{Y.}~\bibnamefont{Wang}} \bibnamefont{and}
  \bibinfo{author}{\bibfnamefont{B.}~\bibnamefont{Yang}},
  \emph{\bibinfo{title}{Geometric fluctuation of conformal hilbert spaces and
  multiple graviton modes in fractional quantum hall effect}}
  (\bibinfo{year}{2022}), \eprint{2201.00020}.

\bibitem[{\citenamefont{Wen}(1991{\natexlab{b}})}]{Wen91b}
\bibinfo{author}{\bibfnamefont{X.}~\bibnamefont{Wen}}, \bibinfo{journal}{Modern
  Physics Letters B} \textbf{\bibinfo{volume}{05}}, \bibinfo{pages}{39}
  (\bibinfo{year}{1991}{\natexlab{b}}),
  \eprint{https://www.worldscientific.com/doi/pdf/10.1142/S0217984991000058},
  \urlprefix\url{https://www.worldscientific.com/doi/abs/10.1142/S0217984991000058}.

\bibitem[{\citenamefont{Wen}(1992)}]{Wen92b}
\bibinfo{author}{\bibfnamefont{X.-G.} \bibnamefont{Wen}},
  \bibinfo{journal}{International Journal of Modern Physics B}
  \textbf{\bibinfo{volume}{06}}, \bibinfo{pages}{1711} (\bibinfo{year}{1992}),
  \urlprefix\url{http://www.worldscientific.com/doi/abs/10.1142/S0217979292000840}.

\bibitem[{\citenamefont{Moore and Wen}(1998)}]{Moore98}
\bibinfo{author}{\bibfnamefont{J.~E.} \bibnamefont{Moore}} \bibnamefont{and}
  \bibinfo{author}{\bibfnamefont{X.-G.} \bibnamefont{Wen}},
  \bibinfo{journal}{Phys. Rev. B} \textbf{\bibinfo{volume}{57}},
  \bibinfo{pages}{10138} (\bibinfo{year}{1998}),
  \urlprefix\url{http://link.aps.org/doi/10.1103/PhysRevB.57.10138}.

\bibitem[{\citenamefont{Banerjee et~al.}(2017)\citenamefont{Banerjee, Heiblum,
  Rosenblatt, Oreg, Feldman, Stern, and Umansky}}]{Banerjee17}
\bibinfo{author}{\bibfnamefont{M.}~\bibnamefont{Banerjee}},
  \bibinfo{author}{\bibfnamefont{M.}~\bibnamefont{Heiblum}},
  \bibinfo{author}{\bibfnamefont{A.}~\bibnamefont{Rosenblatt}},
  \bibinfo{author}{\bibfnamefont{Y.}~\bibnamefont{Oreg}},
  \bibinfo{author}{\bibfnamefont{D.~E.} \bibnamefont{Feldman}},
  \bibinfo{author}{\bibfnamefont{A.}~\bibnamefont{Stern}}, \bibnamefont{and}
  \bibinfo{author}{\bibfnamefont{V.}~\bibnamefont{Umansky}},
  \bibinfo{journal}{Nature} \textbf{\bibinfo{volume}{545}}, \bibinfo{pages}{75}
  (\bibinfo{year}{2017}), ISSN \bibinfo{issn}{0028-0836}.

\bibitem[{\citenamefont{Banerjee et~al.}(2018)\citenamefont{Banerjee, Heiblum,
  Umansky, Feldman, Oreg, and Stern}}]{Banerjee17b}
\bibinfo{author}{\bibfnamefont{M.}~\bibnamefont{Banerjee}},
  \bibinfo{author}{\bibfnamefont{M.}~\bibnamefont{Heiblum}},
  \bibinfo{author}{\bibfnamefont{V.}~\bibnamefont{Umansky}},
  \bibinfo{author}{\bibfnamefont{D.~E.} \bibnamefont{Feldman}},
  \bibinfo{author}{\bibfnamefont{Y.}~\bibnamefont{Oreg}}, \bibnamefont{and}
  \bibinfo{author}{\bibfnamefont{A.}~\bibnamefont{Stern}},
  \bibinfo{journal}{Nature} \textbf{\bibinfo{volume}{559}},
  \bibinfo{pages}{205} (\bibinfo{year}{2018}), ISSN \bibinfo{issn}{1476-4687},
  \urlprefix\url{https://doi.org/10.1038/s41586-018-0184-1}.

\bibitem[{\citenamefont{Srivastav et~al.}(2019)\citenamefont{Srivastav, Sahu,
  Watanabe, Taniguchi, Banerjee, and Das}}]{Srivastav19}
\bibinfo{author}{\bibfnamefont{S.~K.} \bibnamefont{Srivastav}},
  \bibinfo{author}{\bibfnamefont{M.~R.} \bibnamefont{Sahu}},
  \bibinfo{author}{\bibfnamefont{K.}~\bibnamefont{Watanabe}},
  \bibinfo{author}{\bibfnamefont{T.}~\bibnamefont{Taniguchi}},
  \bibinfo{author}{\bibfnamefont{S.}~\bibnamefont{Banerjee}}, \bibnamefont{and}
  \bibinfo{author}{\bibfnamefont{A.}~\bibnamefont{Das}},
  \bibinfo{journal}{Science Advances} \textbf{\bibinfo{volume}{5}}
  (\bibinfo{year}{2019}),
  \eprint{https://advances.sciencemag.org/content/5/7/eaaw5798.full.pdf},
  \urlprefix\url{https://advances.sciencemag.org/content/5/7/eaaw5798}.

\bibitem[{\citenamefont{Srivastav et~al.}(2021)\citenamefont{Srivastav, Kumar,
  Sp\aa{}nsl\"att, Watanabe, Taniguchi, Mirlin, Gefen, and Das}}]{Srivastav21}
\bibinfo{author}{\bibfnamefont{S.~K.} \bibnamefont{Srivastav}},
  \bibinfo{author}{\bibfnamefont{R.}~\bibnamefont{Kumar}},
  \bibinfo{author}{\bibfnamefont{C.}~\bibnamefont{Sp\aa{}nsl\"att}},
  \bibinfo{author}{\bibfnamefont{K.}~\bibnamefont{Watanabe}},
  \bibinfo{author}{\bibfnamefont{T.}~\bibnamefont{Taniguchi}},
  \bibinfo{author}{\bibfnamefont{A.~D.} \bibnamefont{Mirlin}},
  \bibinfo{author}{\bibfnamefont{Y.}~\bibnamefont{Gefen}}, \bibnamefont{and}
  \bibinfo{author}{\bibfnamefont{A.}~\bibnamefont{Das}},
  \bibinfo{journal}{Phys. Rev. Lett.} \textbf{\bibinfo{volume}{126}},
  \bibinfo{pages}{216803} (\bibinfo{year}{2021}),
  \urlprefix\url{https://link.aps.org/doi/10.1103/PhysRevLett.126.216803}.

\bibitem[{\citenamefont{Srivastav et~al.}(2022)\citenamefont{Srivastav, Kumar,
  Spånslätt, Watanabe, Taniguchi, Mirlin, Gefen, and Das}}]{Ravi}
\bibinfo{author}{\bibfnamefont{S.~K.} \bibnamefont{Srivastav}},
  \bibinfo{author}{\bibfnamefont{R.}~\bibnamefont{Kumar}},
  \bibinfo{author}{\bibfnamefont{C.}~\bibnamefont{Spånslätt}},
  \bibinfo{author}{\bibfnamefont{K.}~\bibnamefont{Watanabe}},
  \bibinfo{author}{\bibfnamefont{T.}~\bibnamefont{Taniguchi}},
  \bibinfo{author}{\bibfnamefont{A.~D.} \bibnamefont{Mirlin}},
  \bibinfo{author}{\bibfnamefont{Y.}~\bibnamefont{Gefen}}, \bibnamefont{and}
  \bibinfo{author}{\bibfnamefont{A.}~\bibnamefont{Das}},
  \emph{\bibinfo{title}{Determination of topological edge quantum numbers of
  fractional quantum hall phases}} (\bibinfo{year}{2022}),
  \urlprefix\url{https://arxiv.org/abs/2202.00490}.

\bibitem[{\citenamefont{Read}(2009)}]{Read09}
\bibinfo{author}{\bibfnamefont{N.}~\bibnamefont{Read}}, \bibinfo{journal}{Phys.
  Rev. B} \textbf{\bibinfo{volume}{79}}, \bibinfo{pages}{045308}
  (\bibinfo{year}{2009}),
  \urlprefix\url{http://link.aps.org/doi/10.1103/PhysRevB.79.045308}.

\end{thebibliography}
	\bibliographystyle{apsrev}
\end{document}